\documentclass[11pt]{article}

\usepackage[top=1in,bottom=1in,left=1in,right=1in]{geometry}
\usepackage{natbib}
\usepackage[unicode=true,bookmarks=true,bookmarksnumbered=false,bookmarksopen=false,breaklinks=false,pdfborder={0 0 1},backref=false,colorlinks=true]{hyperref}
\hypersetup{citecolor=blue}
\usepackage{url}

\usepackage{amsmath}
\usepackage{mathrsfs}
\usepackage{amssymb}
\usepackage{amsthm}
\usepackage[normalem]{ulem}

\newtheorem{theorem}{Theorem}

\usepackage{float}
\usepackage{rotfloat}

\floatstyle{ruled}
\newfloat{algorithm}{tpb}{loa}
\providecommand{\algorithmname}{Algorithm}
\floatname{algorithm}{\protect\algorithmname}

\usepackage{algpseudocode,algorithm}
\algnewcommand\algorithmicinput{\textbf{Input}:}
\algnewcommand\algorithmicoutput{\textbf{Output}:}
\algnewcommand\INPUT{\item[\algorithmicinput]}
\algnewcommand\OUTPUT{\item[\algorithmicoutput]}

\usepackage{graphicx}
\usepackage[caption=false,labelformat=simple]{subfig}

\usepackage{array}
\newcolumntype{L}[1]{>{\raggedright\let\newline\\\arraybackslash\hspace{0pt}}m{#1}}
\newcolumntype{C}[1]{>{\centering\let\newline\\\arraybackslash\hspace{0pt}}m{#1}}
\newcolumntype{R}[1]{>{\raggedleft\let\newline\\\arraybackslash\hspace{0pt}}m{#1}}

\usepackage{rotating}
\usepackage{multirow}

\usepackage{tikz}
\usetikzlibrary{shapes,arrows,backgrounds,calc,positioning,fit,petri,plotmarks}
\usetikzlibrary{arrows}

\usepackage{enumerate}
\usepackage{paralist}

\newcommand*{\affaddr}[1]{#1} 
\newcommand*{\affmark}[1][*]{\textsuperscript{#1}}

\global\long\def\bx{\mathbf{x}}
\global\long\def\bY{\mathbf{Y}}
\global\long\def\by{\mathbf{y}}

\global\long\def\bz{\mathbf{z}}

\global\long\def\bA{\mathbf{A}}

\global\long\def\bH{\mathbf{H}}

\global\long\def\bw{\mathbf{w}}

\global\long\def\bV{\mathbf{V}}
\global\long\def\bv{\mathbf{v}}

\global\long\def\bS{\mathbf{S}}

\global\long\def\bbeta{\boldsymbol{\beta}}

\global\long\def\bSigma{\boldsymbol{\Sigma}}
\global\long\def\bgamma{\boldsymbol{\gamma}}
\global\long\def\bdeta{\boldsymbol{\eta}}

\global\long\def\bOmega{\boldsymbol{\Omega}}
\global\long\def\bPi{\boldsymbol{\Pi}}
\global\long\def\bpi{\boldsymbol{\pi}}
\global\long\def\bLambda{\boldsymbol{\Lambda}}

\global\long\def\bGamma{\boldsymbol{\Gamma}}
\global\long\def\bTheta{\boldsymbol{\Theta}}

\global\long\def\balpha{\boldsymbol{\alpha}}

\usepackage{xr}
\makeatletter
\newcommand*{\addFileDependency}[1]{
  \typeout{(#1)}
  \@addtofilelist{#1}
  \IfFileExists{#1}{}{\typeout{No file #1.}}
}
\makeatother
 
\newcommand*{\myexternaldocument}[1]{%
    \externaldocument{#1}%
    \addFileDependency{#1.tex}%
    \addFileDependency{#1.aux}%
}

\myexternaldocument{}

\title{Parsimonious Clustering of Covariance Matrices}

\author{%
    Yixi Xu\affmark[1] and Yi Zhao\affmark[1]
    \\
    \affaddr{\affmark[1]Department of Biostatistics and Health Data Science, Indiana University School of Medicine} \\
}

\date{}

\providecommand{\keywords}[1]
{
  {\small 
  \textbf{Keywords:} #1 Common diagonalization; Covariance regression; Decomposition method; Mixture-of-Expert model}
}

\begin{document}

\maketitle

\thispagestyle{empty}

\begin{abstract}
Functional connectivity (FC) derived from functional magnetic resonance imaging (fMRI) data offers vital insights for understanding brain function and neurological and psychiatric disorders. Unsupervised clustering methods are desired to group individuals based on shared features, facilitating clinical diagnosis. In this study, a parsimonious clustering model is proposed, which integrates the Mixture-of-Experts (MoE) and covariance regression framework, to cluster individuals based on FC captured by data covariance matrices in resting-state fMRI studies. The model assumes common linear projections across covariance matrices and a generalized linear model with covariates, allowing for flexible yet interpretable projection-specific clustering solutions. To evaluate the performance of the proposed framework, extensive simulation studies are conducted to assess clustering accuracy and robustness. The approach is applied to resting-state fMRI data from the Alzheimer's Disease Neuroimaging Initiative (ADNI). Subgroups are identified based on brain coherence and simultaneously uncover the association with demographic factors and cognitive functions.
\end{abstract}



\clearpage
\setcounter{page}{1}

\section{Introduction}
\label{sec:intro}
Functional Magnetic Resonance Imaging (fMRI) is a widely used non-invasive technique for measuring brain activity by detecting fluctuations in the Blood Oxygen Level Dependent (BOLD) signal, which serves as a proxy for neural activity. Functional connectivity (FC), often referred to as the brain functional connectome, is typically derived by measuring the temporal coherence of BOLD signals across predefined brain Regions of Interest (ROIs), each representing a set of voxels based on a brain atlas. One generally used way of quantifying FC is to calculate the covariance matrix of standardized fMRI signals extracted from brain regions. It captures the degree of signal coherence between brain regions, reflecting underlying neural interactions~\citep{friston2011functional}. Recent studies have shown increasing interest in using resting-state fMRI data to cluster subjects based on their FC patterns, aiming to identify clinically meaningful subtypes of psychiatric and neurological disorders, such as schizophrenia, depression, and Alzheimer's disease~\citep{zeng2014unsupervised,dennis2014functional,clementz2016identification,yamada2017resting,yang2022study}. The findings underscore the value of FC-based clustering in uncovering brain network dysfunctions associated with specific disease phenotypes.

Functional MRI data generally feature a common zero-mean structure across groups but exhibit different covariance structures after a centralization processing step~\citep{chen2022simultaneous}. While much of the clustering literature has focused on mean-based methods, relatively less attention has been given to covariance-based clustering. Covariance-based clustering methods can generally be categorized into algorithm-based and model-based approaches. Algorithm-based approaches use similarity metrics to define cluster membership. Examples includes geodesic distance~\citep{lee2015geodesic}, global covariance structure distance~\citep{ieva2016}, and the Wasserstein metric~\citep{verdinelli2019hybrid}. In contrast, model-based clustering approaches provide a framework for simultaneously inferring cluster membership and modeling parameters of the underlying data distributions. First introduced by \citet{banfield1993model}, it assumes that observations arise from a mixture of probability distributions, with each component corresponding to a distinct cluster. \citet{celeux1995gaussian} later extended this work by employing maximum likelihood estimation via the Expectation-Maximization (EM) algorithm~\citep{dempster1977maximum}, with Gaussian Mixture Models (GMMs) being a commonly used example. The EM algorithm estimates parameters and assigns observations to clusters based on posterior probabilities. 
Recent developments, such as the Mixture-of-Experts (MoE) model~\citep{jacobs1991adaptive}, offered greater flexibility by incorporating covariates into the clustering process. This framework has been further extended in approaches like MoEClust~\citep{murphy2020gaussian}, which generalizes model-based clustering by assuming that the cluster assignment depends on a set of covariates of interest. Model-based methods are flexible and particularly suited for neuroimaging applications, as they provide robust solutions for determining the number of clusters and offer a probabilistic interpretation of cluster membership for complex data structures.

Despite these advancements, relatively few studies utilizing model-based clustering have focused on clustering subjects solely based on the covariance matrices derived from fMRI data without prior labels. \citet{dilernia2022penalized} proposed a penalized random covariance clustering approach to simultaneously identify subject clusters and estimate sparse precision matrices for each cluster based on FC patterns. Similarly, \citet{liu2023simultaneous} introduced a framework for clustering and estimating heterogeneous graphs from matrix-valued fMRI data. Recently, \citet{lin2024latent} developed an image-on-scalar regression model for heterogeneous image data to identify clusters and explore their association with an exposure variable of interest.
Though \citet{dilernia2022penalized} and \citet{liu2023simultaneous} cluster subjects based on covariance or precision matrices, the approaches are fully unsupervised and do not utilize covariate information to guide clustering. As a result,  these methods cannot assess how subject-level factors are related to identified clusters. \citet{lin2024latent} adopted a semi-supervised formulation that incorporates covariates, but its framework is designed for voxel- or image-level data rather than covariance matrices derived for FC.

In this paper, we propose a parsimonious clustering model (\textbf{CAPclust}) that integrates the MoE framework with covariance regression~\citep{zhao2021covariate} to cluster subjects based on covariance matrices. This approach is motivated by resting-state fMRI studies, where brain FC is captured by the covariance matrix of fMRI signals. The proposed framework enhances flexibility and interpretability by incorporating covariate effects through generalized linear models (GLMs) and ensuring the positive definiteness of covariance matrices. By assuming common linear projections on the covariance matrices, the proposed approach interprets these as distinct brain subnetworks and enables subnetwork-specific clustering, facilitating a deeper understanding of how covariates influence brain connectivity patterns and individual clustering. This framework allows for a better understanding of the relationship between network structure and subject-specific covariates, contributing to the interpretation of clustering results in fMRI studies.

The primary contributions of this paper are the following.  
\begin{inparaenum}[(i)\upshape]
\item 
A novel covariance clustering approach is introduced, which incorporates covariance regression and the MoE modeling framework.
\item 
Unlike other existing approaches, the proposed approach allows component-specific clustering, enabling distinct clusters to be identified for different dimensions (or projections) of the covariance matrices.
\item The proposed framework offers a better understanding of brain network subgrouping in resting-state fMRI studies. 
\end{inparaenum}

The remainder of the paper is organized as follows. Section~\ref{sec:meth} presents the proposed covariance matrix based clustering model, estimation method, asymptotic properties, and computational algorithm, along with the procedure for selecting the number of clusters. In Section~\ref{sec:sim}, simulation studies are conducted to evaluate the performance of the proposed approach in comparison to existing approaches. Section~\ref{sec:application} demonstrates an application to the data from the Alzheimer's Disease Neuroimaging Initiative (ADNI). Finally, Section~\ref{sec:discussion} concludes the manuscript with discussions.

\section{Methods}
\label{sec:meth}
Let $\by_{it}\in \mathbb{R}^{p}$ denote a $p$-dimensional random vector for subject $i$ and observation $t$, for $i=1,\dots,n$ and $t = 1, \dots, T_{i}$, where $T_{i}$ is the number of observations of the $i$th subject and $n$ is the total number of subjects. 
In an fMRI study, $\by_{it}$ represents the $t$th volume of the scan collected from subject $i$.
Assume $\by_{it}$'s, for $t=1,\dots,T_{i}$, are independently and identically distributed with mean of zero and covariance matrix $\bSigma_{i}$. Here, as the study interest focuses on the covariance matrix, without loss of generality, it is assumed that the distribution mean is zero. In practice, one can achieve this by centralizing the data. For fMRI studies, one may also scale the data to unit variance so that the covariance and correlation matrices are identical.
For the covariance matrix of subject $i$, $\bSigma_{i}$, it is assumed that there exists a linear projection,  $\bgamma \in \mathbb{R}^{p}$, which is shared across subjects, such that $z_{it} = \bgamma^{\top} \by_{it}$ follows a Gaussian mixture distribution with $K$ members:
\begin{equation}\label{eq:model_mix}
f(z_{it}\mid  \pi_{ik},\sigma_{i k}^2) = \sum_{k=1}^K \pi_{i k} \mathcal{N}(0, \sigma_{i k}^2),
\end{equation}
where $\pi_{ik}$ is the mixing proportion such that $0 \leq \pi_{i k} \leq 1$ and  $\sum_{k=1}^K \pi_{i k}=1$, for $i=1, \ldots, n$. $\pi_{i k}$ can be interpreted as the probability of subject $i$ belonging to cluster $k$. The member density, $\mathcal{N}(0, \sigma_{i k}^2)$, or known as the expert network, is a Gaussian distribution with mean zero and variance $\sigma_{i k}^2$. Here, it is assumed that $\sigma_{ik}^{2}$ depends on a set of covariates, $\bx_{i} \in \mathbb{R}^{q_{1}}$, such that
\begin{equation}\label{eq:model_var}
\log \left(\sigma_{i k}^2\right)=\bx_i^{\top} \bbeta_k, \text{ or } \sigma_{i k}^2=\exp \left(\bx_i^{\top} \bbeta_k\right),
\end{equation}
where $\bbeta_{k}$ is the model coefficient for the $k$th cluster, for $k=1,\dots,K$.
In the context of MoE, the mixing probabilities are modeled as a function of a set of covariates, $\bw_{i} \in \mathbb{R}^{q_{2}}$. Here, to make it more general, $\bx_{i}$ and $\bw_{i}$ can be two different sets of covariates. These mixing probabilities, also known as the gating functions, are modeled by a multinomial logistic regression and defined as
\begin{equation}\label{eq:model_pi}
\pi_{i k}=\pi_k\left(\mathbf{w}_i\right)=\frac{\exp \left(\mathbf{w}_i^{\top} \boldsymbol{\alpha}_k\right)}{\sum_{k=1}^K \exp \left(\mathbf{w}_i^{\top} \boldsymbol{\alpha}_k\right)}, \quad \text{for } k=2,\dots,K,
\end{equation}
where $\balpha_{k}$ is the clustering model coefficient. For $k=1$, we set $\balpha_1 = (0,\ldots,0)^\top = \boldsymbol{\mathrm{0}}$ for identification. 
This semi-supervised formulation allows the clustering to be guided by subject-level covariates, thereby improving interpretability and stabilizing estimation. It enables prediction of cluster membership for new subjects given their covariates, extending the framework beyond unsupervised covariance-based clustering.

\subsection{Estimation}
\label{sub:estimation}

Different from unsupervised clustering, the proposed framework enables covariate-guided clustering through parametric regression models. To estimate these parameters, where the parameter space is denoted as $\bTheta = {\left\{\boldsymbol{\gamma},\left\{\boldsymbol{\beta}_1, \ldots, \boldsymbol{\beta}_K\right\},\left\{\boldsymbol{\alpha}_1, \ldots, \boldsymbol{\alpha}_K\right\}\right\}}$, together with simultaneous cluster allocation, a likelihood-based approach is considered. The full mixture likelihood is written as
 \begin{equation}
  \mathcal{L}=\prod_{i=1}^n \prod_{t=1}^{T_i} \prod_{k=1}^K\left\{f_k\left(z_{i t} \mid \mathbf{x}_i\right) \pi_k\left(\mathbf{w}_i\right)\right\}^{\mathrm{I}\left(\eta_i=k\right)}, \\
 \end{equation} 
where $f_k\left(z_{i t} \mid \mathbf{x}_i\right)$ is the density of normal distribution with mean zero and variance depending on $\bx_{i}$, $\pi_k\left(\mathbf{w}_i\right)$ is defined above by~\eqref{eq:model_pi}, $\eta_{i}$ is a latent indicator of the cluster membership, and $\mathrm{I}(\cdot)$ is an indicator function.
Let $\bdeta_{i}=(\eta_{i1},\dots,\eta_{iK})^\top\in\mathbb{R}^{K}$ be a vector of cluster membership indicators, where $\eta_{ik}=\mathrm{I}(\eta_{i}=k)$ is one if subject $i$ is a member from cluster $k$ and zero otherwise.
The complete data log-likelihood is
\begin{eqnarray}\label{eq:loglik_complete}
\ell\left(\bTheta \mid \mathbf{y}_{i t}, \mathbf{x}_i, \mathbf{w}_i, \boldsymbol{\eta}_i\right) 
=\sum_{i=1}^n \sum_{k=1}^K \eta_{i k} \frac{T_i}{2}\left\{2 \log \pi_k\left(\mathbf{w}_i\right)-\mathbf{x}_i^{\top} \boldsymbol{\beta}_k-\exp \left(-\mathbf{x}_i^{\top} \boldsymbol{\beta}_k\right)\left(\boldsymbol{\gamma}^{\top} \mathbf{S}_i \boldsymbol{\gamma}\right)\right\} \nonumber \\
=\sum_{i=1}^n \sum_{k=1}^K T_i \eta_{i k} \log \pi_k\left(\mathbf{w}_i\right)+\sum_{i=1}^n \sum_{k=1}^K \frac{T_i}{2} \eta_{i k}\left\{-\mathbf{x}_i^{\top} \boldsymbol{\beta}_k-\exp \left(-\mathbf{x}_i^{\top} \boldsymbol{\beta}_k\right)\left(\boldsymbol{\gamma}^{\top} \mathbf{S}_i \boldsymbol{\gamma}\right)\right\}, 
\end{eqnarray}
\sloppy
where $\bS_{i}=\sum_{t=1}^{T_{i}}\by_{it}\by_{it}^\top/T_{i}$ is the sample covariance matrix of subject $i$.
With latent indicators, a popular way of estimating the parameters is to utilize the Expectation-Maximization (EM) algorithm.

\subsubsection{E-step}
\label{sub:E-step}

For the E-step, with the estimate from the $s$th step, $\hat{\bTheta}^{(s)}$, for $s=0,1,2,\dots$, we first define 
\begin{equation} \label{eq:compute_eta}
\hat{\eta}_{i k}^{(s+1)}=\mathbb{E}\left(\eta_{i k} \mid \mathbf{y}_{i t}, \mathbf{x}_i, \mathbf{w}_i, \hat{\bTheta}^{(s)}\right)
=\frac{\hat{\pi}_k^{(s)}\left(\mathbf{w}_i\right) \hat{\phi}_k^{(s)}\left(\mathbf{z}_i \mid \mathbf{x}_i\right)}{\sum_{k=1}^K \hat{\pi}_k^{(s)}\left(\mathbf{w}_i\right) \hat{\phi}_k^{(s)}\left(\mathbf{z}_i \mid \mathbf{x}_i\right)}.
\end{equation}
The function $\phi_k$ is the probability density function (PDF) of a normal distribution, and $\hat{\phi}_k$ denotes the estimated density by plugging in the estimated parameters. The conditional expectation of the complete objective function is 
    \begin{align}
    & Q(\bTheta,\bTheta^{(s)})  \\ 
    = & \mathbb{E}\left\{\ell\left(\bTheta \mid \mathbf{y}_{i t}, \mathbf{x}_i, \mathbf{w}_i, \boldsymbol{\eta}_i, \hat{\bTheta}^{(s)}\right)\right\} \nonumber \\
    =  & \sum_{i=1}^n \sum_{k=1}^K T_i \hat{\eta}_{i k}^{(s+1)} \frac{\exp \left(\mathbf{w}_i^{\top} \boldsymbol{\alpha}_k\right)}{\sum_{k=1}^K \exp \left(\mathbf{w}_i^{\top} \boldsymbol{\alpha}_k\right)} +\sum_{i=1}^n \sum_{k=1}^K \frac{T_i}{2} \hat{\eta}_{i k}^{(s+1)}\left\{-\mathbf{x}_i^{\top} \boldsymbol{\beta}_k-\exp \left(-\mathbf{x}_i^{\top} \boldsymbol{\beta}_k\right)\left(\boldsymbol{\gamma}^{\top} \mathbf{S}_i \boldsymbol{\gamma}\right)\right\}. \nonumber
    \end{align}

\subsubsection{M-step}
\label{sub:M-step}

For the M-step, to maximize $Q(\bTheta,\bTheta^{(s)})$, a block-coordinate descent algorithm is considered to optimize over each parameter separately.
For $\balpha_k$, the first term of the equation can be viewed as the multinomial logistic regression with weights $T_i$ and covariates $\bw_i$. The solution can be obtained by fitting the regression model. For $\bbeta_k$, it is equivalent to minimizing the following objective function, 
\begin{equation}
\ell_{s+1}=\sum_{i=1}^n \sum_{k=1}^K \frac{T_i}{2} \hat{\eta}_{i k}^{(s+1)}\left\{\mathbf{x}_i^{\top} \boldsymbol{\beta}_k+\exp \left(-\mathbf{x}_i^{\top} \boldsymbol{\beta}_k\right)\left(\boldsymbol{\gamma}^{\top} \mathbf{S}_i \boldsymbol{\gamma}\right)\right\}.
\end{equation}
The Newton-Raphson method is employed to obtain the solution. The update is 
\begin{equation}
    \hat{\boldsymbol{\beta}}_k^{(s+1)}=\hat{\boldsymbol{\beta}}_k^{(s)}-\left.\left.\left(\frac{\partial^2 \ell_{s+1}}{\partial \boldsymbol{\beta}_k \partial \boldsymbol{\beta}_k^{\top}}\right)^{-1}\right|_{\boldsymbol{\beta}_k^{(s)}, \boldsymbol{\gamma}^{(s)}}\left(\frac{\partial \ell_{s+1}}{\partial \boldsymbol{\beta}_k}\right)\right|_{\boldsymbol{\beta}_k^{(s)}, \boldsymbol{\gamma}^{(s)}},
\end{equation}
where
\begin{eqnarray}
    \frac{\partial \ell_{s+1}}{\partial \boldsymbol{\beta}_k} &=& \sum_{i=1}^n \frac{T_i}{2} \hat{\eta}_{i k}^{(s+1)}\left\{1-\exp \left(-\mathbf{x}_i^{\top} \hat{\bbeta}^{(s)}_k\right)\left(\hat{\bgamma}^{\top(s)} \mathbf{S}_i \hat{\bgamma}^{(s)}\right)\right\} \mathbf{x}_i, \\
\frac{\partial^2 \ell_{s+1}}{\partial \boldsymbol{\beta}_k \partial \boldsymbol{\beta}_k^{\top}} &=& \sum_{i=1}^n \frac{T_i}{2} \hat{\eta}_{i k}^{(s+1)} \exp \left(-\mathbf{x}_i^{\top} \hat{\bbeta}^{(s)}_k\right)\left(\hat{\bgamma}^{\top(s)} \mathbf{S}_i \hat{\bgamma}^{(s)}\right) \mathbf{x}_i \mathbf{x}_i^{\top},
\end{eqnarray}
$\hat{\boldsymbol{\beta}}_k^{(s)}$ and $\hat{\bgamma}^{(s)}$ are the solutions from the $s$th iteration, and $\hat{\eta}_{i k}^{(s+1)}$ is defined above in~\eqref{eq:compute_eta}.
For $\bgamma$, it is proposed to solve the following optimization problem with a constraint for identification.
\begin{eqnarray}\label{eq:solve_gamma}
  \text { minimize } && \tilde{\ell}_{s+1}=\boldsymbol{\gamma}^{\top}\left(\sum_{i=1}^n \sum_{k=1}^K \frac{T_i}{2} \hat{\eta}_{i k}^{(s+1)} \exp \left(-\mathbf{x}_i^{\top} \hat{\bbeta}^{(s+1)}_k\right) \mathbf{S}_i\right) \boldsymbol{\gamma}, \nonumber \\
    \text{ such that } && \boldsymbol{\gamma}^{\top} \mathbf{H} \boldsymbol{\gamma}=1,
\end{eqnarray}
\sloppy where $\bH$ is a $p \times p$ positive definite matrix. For a likelihood-based approach, one can choose $\bH$ as $\bH=\bar{\bS}=\sum_{i=1}^{n}\sum_{j=1}^{T_{i}}\by_{it}\by_{it}^\top/\sum_{i=1}^{n}T_{i}$, the average of the sample covariance matrices, to incorporate distribution information of the data into optimization. This type of choice is also considered in~\citet{zhao2021covariate}. See the supplementary materials of \citet{zhao2021covariate} for a detailed discussion about the choice.
To simplify the notation, let $\mathbf{A}^{(s+1)}= \sum_{i=1}^n \sum_{k=1}^K (T_{i}/2) \hat{\eta}_{i k}^{(s+1)} \exp \left(-\mathbf{x}_i^{\top} \hat{\boldsymbol{\beta}}_k^{(s+1)}\right) \mathbf{S}_i$. It can be shown that it is equivalent to solving the following
\begin{equation}
\mathbf{A}^{(s+1)} \boldsymbol{\gamma} - \lambda \bH \bgamma = 0,
\end{equation}
where $\lambda$ is the Lagrangian multiplier. The solution is the eigenvector and corresponding eigenvalue of $\bA^{(s+1)}$  with respect to $\bH$. Details are available in Section~\ref{appendix:sec:solve_gamma} of the supplementary materials. Algorithm~\ref{alg:EM} summarizes the estimation procedure using the EM.

\begin{algorithm}
    \caption{\label{alg:EM}An Expectation-Maximization algorithm of solving~\eqref{eq:loglik_complete}.}
    \begin{algorithmic}[1]
        \INPUT $\{\{\by_{it}\}_{t},\bx_{i},\bw_{i}\}_{i}$
        \State \textbf{Initialization}:
        $(\bgamma^{(0)},\{\balpha_{k}^{(0)},\bbeta_{k}^{(0)}\}_{k})$
        
        \Repeat \; for iteration $s=0,1,2,\dots$,

         \State \; update \(\eta_{ik}\) with~\eqref{eq:compute_eta}, denoted as \(\hat{\eta}_{ik}^{(s+1)}\), for $i=1,\dots,n$ and $k=1,\dots,K$,

            \State \; update $ \balpha_{k}$ using the  multinomial logistic regression, denoted as $\hat{\balpha}_{k}^{(s+1)}$, for $k=2,\dots,K$,

             \State \; update $ \bbeta_{k}$ using the Newton-Raphson algorithm, denoted as $\hat{\bbeta}_{k}^{(s+1)}$, for $k=1,\dots,K$,

             \State \; update $\bgamma$ by solving~\eqref{eq:solve_gamma}, denoted as $\hat{\bgamma}^{(s+1)}$,
      
        \Until{the objective function in~\eqref{eq:loglik_complete} converges;}

        \State Consider a random series of initializations, repeat Steps 1--7, and choose the solution with the minimum objective value.

        \OUTPUT $(\hat{\bgamma},\{\hat{\balpha}_{k},\hat{\bbeta}_{k}\}_{k},\{\hat{\eta}_{ik}\}_{i,k})$
    \end{algorithmic}
\end{algorithm}

\subsection{Higher-order components and choose the number of components}

Section~\ref{sub:estimation} discusses the parameter identification and clustering of one projecting component.
For higher-order components, it is proposed to identify them sequentially, analogous to the approach in principal component analysis that identifies orthogonal components. 
This allows the clustering to differ across components.
Let $\boldsymbol{\Gamma}^{(r-1)}=\left(\bgamma_1, \ldots, \bgamma_{r-1}\right) \in \mathbb{R}^{p \times(r-1)}$ denote the first $(r-1)$ identified components. 
Let $\bY_{i}=(\by_{i1},\dots,\by_{iT_{i}})^\top\in\mathbb{R}^{T_{i}\times p}$, for $i=1,\dots,n$.
To remove the identified components, set 
\begin{equation}
\hat{\mathbf{Y}}_i^{(r)}=\mathbf{Y}_i-\mathbf{Y}_i \boldsymbol{\Gamma}^{(r-1)} \boldsymbol{\Gamma}^{(r-1) \top}.
\end{equation}
One can show that $\hat{\bY}_{i}^{(r)}$ lies in the space orthogonal to the space spanned by $\bGamma^{(r-1)}$.
Since $\hat{\mathbf{Y}}_i^{(r)}$ is not of full rank, one should add an additional rank-completion step following the algorithm introduced in~\citet{zhao2021covariate}. With the rank-completed data as the input of Algorithm~\ref{alg:EM}, it identifies the next component that is orthogonal to $\bGamma^{(r-1)}$.

To determine the number of components, the following metric, called average deviation from diagonality (DfD) introduced in \citet{zhao2021covariate}, is employed.
\begin{equation}\label{eq:DfD}
  \mathrm{DfD}(\hat{\bGamma}^{(r)})=\prod_{i=1}^{n}\left[\frac{\det\left\{\mathrm{diag}(\hat{\bGamma}^{(r)\top}\bS_{i}\hat{\bGamma}^{(r)})\right\}}{\det(\hat{\bGamma}^{(r)\top}\bS_{i}\hat{\bGamma}^{(r)})}\right]^{T_{i}/\sum_{i}T_{i}},
\end{equation}
where $\mathrm{diag}(\bA)$ is a diagonal matrix taking the diagonal elements in a square matrix $\bA$ and $\det(\bA)$ is the determinant of $\bA$. When $\hat{\bGamma}^{(r)}$ commonly diagonalizes $\bS_{i}$ for all $i\in\{1,\dots,n\}$, the above DfD achieves its minimum value of one. As $r$ increases, it becomes more challenging to diagonalize all estimated covariance matrices, and the value of the $\mathrm{DfD}$ metric may increase significantly. In practice, the number of components can be chosen by setting a threshold value of the DfD or before a sudden jump in the value. Here we set a threshold and choose $r$ as the maximum value that $\mathrm{DfD} \leq 2$ holds, as suggested by \citet{zhao2021covariate}.

\subsection{Inference}
\label{sub:inference}
In the current framework, we focus on the inference of model coefficients, $\{\balpha_{k},\bbeta_{k}\}$, and fix $\bgamma$ as the estimate from the full dataset. 
The following nonparametric bootstrap procedure is considered.
\begin{description}
  \item[Step 0.] Use all sample units and identify $r$ components, where the estimated linear projections are denoted as $\hat{\bgamma}_{j}$, for $j=1,\dots,r$.
  \item[Step 1.] Generate a bootstrap sample of size $n$ by sampling with replacement, denoted as \\ $\{\bY_{i}^{*},\bx_{i}^{*},\bw_{i}^{*} \mid i=1,\dots,n\}$.
  \item[Step 2.] For $j=1,\dots,r$, using the resampled data, estimate model coefficients and variances following the proposed EM Algorithm~\ref{alg:EM} with $\bgamma=\hat{\bgamma}_{j}$.
  \item[Step 3.] Repeat Steps 1--2 for $B$ times.
  \item[Step 4.] Construct bootstrap confidence intervals at a prespecified significance level. 
\end{description}

\subsection{Number of clusters}
\label{sub:ncluster}

To choose the number of clusters, $K$, the Bayesian Information Criterion (BIC) is considered as proposed by \citet{schwarz1978estimating}. The BIC is defined as $\text{BIC}(K) = M \log(\sum_{i}T_{i}) - 2\ell(\hat{\bTheta})$, where $M = Kq_1 + (K-1)q_2 + p$ represents the total number of unknown parameters. 
The optimal number of clusters is determined by minimizing the BIC criterion. 

In our proposed model, we report separate BIC values for each component, as our method allows for suggesting distinct cluster profiles for different components of the covariance matrices. To determine the optimal number of clusters, we use the average BIC over identified components as the selection criterion.

\subsection{Asymptotic properties}
\label{sub:asymp}

This section discusses the asymptotic properties of the proposed approach. We start with the case when the linear projection, $\bgamma$, is known.
\begin{theorem}\label{thm:asymp_gammaknown}
    With $\bgamma$ known, denote $z_{it}=\bgamma^\top\by_{it}$ and $\bz_{i}=(z_{i1},\dots,z_{iT_{i}})\top\in\mathbb{R}^{T_{i}}$. Let $f(\bv;\tilde{\Theta})$ be the density function of $\bV_{i}=(\bx_{i},\bz_{i},\bw_{i})$  with parameter $\tilde{\bTheta}=\bTheta\setminus\{\bgamma\}$. Assume that $f(\bv;\tilde{\bTheta})$ satisfies the regularity conditions R1--R5 in the supplementary materials. Let $\tilde{\ell}(\tilde{\bTheta})$ denote the pseudo-likelihood function with given $\bgamma$. Then, there exists a maximizer $\hat{\tilde{\bTheta}}$ of $\tilde{\ell}(\tilde{\bTheta})$ such that $\|\hat{\tilde{\bTheta}}-\tilde{\bTheta}^{*}\|=\mathcal{O}_{p}(n^{-1/2})$, where $\tilde{\bTheta}^{*}$ is the true parameter.
\end{theorem}
The regularity conditions and proof of the theorem are presented in Section~\ref{appendix:sec:proof} and Section~\ref{sup:asymp_gammaknown} of the supplementary materials, respectively. These conditions are also considered in \citet{khalili2010new} for establishing consistency and oracle properties of penalized likelihood estimators in mixture-of-experts regression models. Theorem~\ref{thm:asymp_gammaknown} suggests that with $\bgamma$ known, the proposed estimator of model coefficients achieves the parametric  $\sqrt{n}$-consistency. To further study the consistency when $\bgamma$ is unknown, additional regularity conditions on the covariance matrices are required.
Assume that the covariance matrix, $\bSigma_{i}$, has the eigendecomposition of $\bSigma_{i}=\bPi_{i}\bLambda_{i}\bPi_{i}^\top$, where $\bPi_{i}=(\bpi_{i1},\dots,\bpi_{ip})\in\mathbb{R}^{p\times p}$ is an orthonormal matrix and $\bLambda_{i}=\mathrm{diag}\{\lambda_{i1},\dots,\lambda_{ip}\}\in\mathbb{R}^{p\times p}$ is a diagonal matrix of corresponding eigenvalues. 
\begin{description}
    \item[Assumption A1] Let $T=\min_{i}T_{i}$. $p\ll T$ is fixed. In addition, $q_{1},q_{2}\ll n$ are fixed.
    \item[Assumption A2] The covariance matrices share the same set of eigenvectors, that is $\bPi_{i}=\bPi$, for $i=1,\dots,n$.
    \item[Assumption A3] For $\forall~i=1,\dots,n$, there exists (at least) one column in $\bPi_{i}$ indexed by $j_{i}$, such that $\bgamma=\bpi_{ij_{i}}$ and models~\eqref{eq:model_mix}--\eqref{eq:model_pi} are satisfied.
\end{description}
Assumption A1 assumes a low-dimensional scenario for both the Gaussian mixture model and the regression models. Under this assumption, the sample covariance matrices are well-conditioned, so as the data log-likelihood function~\eqref{eq:loglik_complete}. Assumption A2 assumes all covariance matrices with identical eigenvectors, but the order of the corresponding eigenvalues may vary. Assumption A3 assumes that the parametric models are correctly specified. Together with conditions R1--R5, the following theorem gives the consistency of the proposed estimator. Details of the proof are in Section~\ref{sup:asymp_gammaunknown} of the supplementary materials.
\begin{theorem}\label{thm:asymp_gammaunknown}
    Assume conditions in Theorem~\ref{thm:asymp_gammaknown} hold. Additionally, with Assumptions A1-A3 satisfied, as $n,T\rightarrow\infty$, the proposed estimator, $\hat{\bTheta}$, is consistent.
\end{theorem}


\section{Simulation Study}
\label{sec:sim}
In this section, we assess the performance of the proposed clustering method, \textbf{CAPclust}, through simulations.  Three simulation studies are conducted: (i) to assess the performance of the CAPclust approach across different sample sizes; (ii) to evaluate the robustness of the method under model misspecification by introducing an additional interaction term in the variance model~\eqref{eq:model_var}; and (iii) to further examine robustness by adding an interaction term to both the variance and mix proportion model~\eqref{eq:model_pi}. 
Additionally, we compare clustering performance with competing methods. In the comparison, we consider two cases with one and two dimensions of the covariance matrices satisfying the mixture model (with different class assignments), while the competing methods use the whole covariance matrix for clustering.
Since no existing method can directly identify distinct cluster patterns for different dimensions of the covariance matrices, we consider the following two adaptations when implementing existing clustering approaches. 1) We compute the sample correlation matrices and vectorize the lower-triangular part (without diagonal elements) after a Fisher $z$-transformation as the input for all competing methods. 2) We compute the value of $\log(\bgamma^\top \bS_{i} \bgamma)$ with the true $\bgamma$ and treat it as the input of K-means, hierarchical, and MoEClust clustering methods. For MoEClust, the R package \textsf{MoEClust} is implemented~\citep{murphy2020gaussian}. A latent subgroup image-on-scalar regression model~\cite[LASIR,][]{lin2024latent} is also considered as a competing method.  With a dimension of $p$, the number of correlation pairs is $p(p-1)/2$. The LASIR method is applied to the Fisher $z$-transformed correlations.

In the simulations, a common eigenstructure is assumed across covariance matrices.
Specifically, the covariance matrix of subject $i$ has the decomposition of $\bSigma_{i} = \bPi \bLambda_{i} \bPi^\top$, where $\bPi = (\bpi_{1}, \dots, \bpi_{p}) \in \mathbb{R}^{p \times p}$ is an orthonormal matrix and \(\bLambda_{i} = \mathrm{diag}(\lambda_{i1}, \dots, \lambda_{ip}) \in \mathbb{R}^{p \times p}\) is a diagonal matrix of eigenvalues. In $\bLambda_{i}$, for the second (D2) and fourth (D4) dimensions, we generate eigenvalues from the log-linear model~\eqref{eq:model_var}. For the remaining dimensions, the eigenvalues are generated from a normal distribution with mean value exponentially decaying from $3$ to $-1$ and standard deviation $0.2$.
For covariate $\bx_{i}$, $q_{1}=3$ is considered and denote $\bx_{i}=(1,x_{i1},x_{i2})^\top$, where $x_{i1}$ is generated from a Bernoulli distribution with probability 0.5 of being one, and $x_{i2}$ is generated from a normal distribution with mean zero and standard deviation one. 
For covariate $\bw_{i}$, a case of $q_{2}=2$ is considered, where $\bw_{i}=(1,w_{i1})$ and $w_{i1}$ is generated from a Bernoulli distribution with probability $0.5$ of being one.
The simulated data, $\by_{it}$, are generated from a mixture of \(K = 2\) members of normal distribution with zero mean and variance \(\sigma_{ik}^2\). The parameter~$\balpha_{2}$ of the mixing portion are set as $(0.5, -1)^\top$ on D2 and $(-0.25, 0.5)^\top$ on D4. The parameter $\bbeta_{k}$'s in the variance model are set to be $\bbeta_{1} = (1,1,-1)^\top$ and $\bbeta_{2} = (-1,-1,1)^\top$ for D2; and $\bbeta_{1} = (0.5,0.5,-0.5)^\top$ and $\bbeta_{2} = (0.5,-0.5,0.5)^\top$ for D4.
Simulations are conducted for $200$ replications with data dimension $p = 50$, sample sizes $n = 50, 100, 500$, and each unit containing $T_{i}=T = 100$ observations.
To evaluate the performance of estimating the projection, \(\bgamma\), we calculate the absolute value of the inner product between the estimated and true projection vectors, denoted as \(|\langle \hat{\bgamma}, \bpi_{j}\rangle|\) for \(j = 2, 4\). This metric, ranging between zero and one, measures the angle between the two vectors, where a value of one suggests identical (up to sign flipping) and a value of zero suggests orthogonal.

Table~\ref{table:sim_est} presents the performance of estimating the projection ($\bgamma$) and corresponding coefficient for variances ($\bbeta$) with $K=2$. The results of estimating \(\balpha\) are omitted from the table as the estimate depends on the selected reference cluster. Simulation (i) examines the performance of the proposed method under different sample sizes.
As the number of units increases, the accuracy of estimating both $\bgamma$ and the corresponding coefficient $\bbeta$ improves. The similarity measure, \(|\langle \hat{\bgamma}, \bpi_{j}\rangle|\), converges to one, indicating an increasing accuracy in estimating the true projection as the sample size grows. The bias and mean squared error (MSE) of estimating $\bbeta$ decrease as the sample size increases. 
Section~\ref{sup:partial_common} of the supplementary materials evaluates the performance of CAPclust to the violation of Assumption A2, where the common eigenstructure assumption is relaxed to partial common eigenstructure. The results show that CAPclust is robust to this relaxation in terms of correctly identifying the projection components. Section~\ref{sup:non_guassian} demonstrates that CAPclust is robust to non-Gaussian distributed data.

Simulations (ii) and (iii) assess the robustness to model misspecification. In Simulation (ii), we introduce an additional interaction term in the variance model~\eqref{eq:model_var}, and in Simulation (iii), an additional interaction term is added to both the variance and clustering models. When implementing the proposed approach, the interaction terms are not included in the model fitting. As shown in Table~\ref{table:sim_est}, under misspecification, the proposed method can still successfully identify the projections under both scenarios, but with bias in estimating the model coefficients as expected. This result suggests the robustness of our approach to model misspecification when estimating the projections.

\begin{table}
  \caption{\label{table:sim_est}Performance of estimating the projection ($\bgamma$) and corresponding coefficient ($\bbeta$) over $200$ replications in the simulation study. SE: standard error; MSE: mean squared error.}
  \begin{center}
   \resizebox{\textwidth}{!}{
    \begin{tabular}{c c l r c r r c r r c r r c r r}
        \hline
        & & & \multicolumn{1}{c}{$\hat{\bgamma}$} && \multicolumn{5}{c}{$\hat{\bbeta}_{1}$} && \multicolumn{5}{c}{$\hat{\bbeta}_{2}$} \\
        \cline{4-4}\cline{6-10}\cline{12-16}
        \multicolumn{1}{c}{Simulation} & \multicolumn{1}{c}{$(p,n,T)$} & & & & \multicolumn{2}{c}{Cluster 1} && \multicolumn{2}{c}{Cluster 2} && \multicolumn{2}{c}{Cluster 1} && \multicolumn{2}{c}{Cluster 2} \\
        \cline{6-7}\cline{9-10}\cline{12-13}\cline{15-16}
        & & & \multicolumn{1}{c}{\multirow{-2}{*}{$|\langle\hat{\bgamma},\bpi_{j}\rangle|$ (SE)}} && \multicolumn{1}{c}{Bias} & \multicolumn{1}{c}{MSE} && \multicolumn{1}{c}{Bias} & \multicolumn{1}{c}{MSE} && \multicolumn{1}{c}{Bias} & \multicolumn{1}{c}{MSE} && \multicolumn{1}{c}{Bias} & \multicolumn{1}{c}{MSE} \\
        \hline
      & &  D2 & $0.983$ ($0.013$) &&$-0.080$ & $0.211$ && $-0.383$ &$1.910$ && $0.013$ & $0.004$ &&$0.033$ & $0.116$ \\
      & \multirow{-2}{*}{$(50,50,100)$} \multirow{-2}{*} & D4 & $0.831$ ($0.055$) && $-0.131$ &$ 0.104$ && $0.161$ & $0.088$ && $0.097$ & $0.042$ && $-0.133$ & $0.068$\\
      \cline{3-16}
       & & D2 & $0.994$ ($0.002$) && $0.005$ & $0.061$ && $-0.062$ & $0.144$ && $0.001$ & $0.001$ && $0.003$ & $0.009$ \\
      \multirow{-2}{*}{(1)} &  \multirow{-2}{*}{$(50,100,100)$}  \multirow{-2}{*} & D4 & $0.919$ ($0.121$) &&$-0.080$& $0.033$ && $0.066$ & $0.040$ && $0.073$ &$0.026$ && $-0.086 $ & $0.040$\\
      \cline{3-16}
       & & D2 & $0.999$ ($0.000$) &&$0.002$ & $0.000$ && $-0.002$ & $0.000$ && $-0.001$ & $0.000$ &&$0.002$ &$0.000$ \\
      & \multirow{-2}{*}{$(50,500,100)$}  \multirow{-2}{*} & D4 & $0.980$ ($0.004$) && $-0.033$ & $ 0.014$ && $0.031$ & $0.012$ && $ 0.039$ &$0.017$ && $-0.033$ &$0.012$\\ 
       \hline        
      & & D2 & $0.988$ ($0.001$) && $0.216$ & $0.238$ && $-0.212$ & $8.078$ && $-0.545$ & $0.372$ && $-0.585$ & $0.570$ 
      \\
       \multirow{-2}{*}{(2)}  & \multirow{-2}{*}{$(50,100,100)$}  \multirow{-2}{*} & D4 & $0.919$ ($0.020$) && $-0.048$ & $0.023$ && $0.204$ & $0.078$ && $0.080$ & $0.064$ && $-0.283$ &$0.112$\\
      \cline{3-16}
       & & D2 & $0.989$ ($0.010$) && $0.182$ & $0.142$ && $-0.313$ & $56.04$ &&$-0.550$ & $0.387$ && $-0.593$ & $0.643$ 
\\
      \multirow{-2}{*}{(3)} & \multirow{-2}{*}{$(50,100,100)$} \multirow{-2}{*} & D4 & $0.917$ ($0.018$) && $0.003$ &$0.016$ &&$0.175$ &$0.071$ && $0.038$ & $0.054$ && $-0.230$ & $0.080$ \\
      \hline
    \end{tabular}
    }
     \end{center}
\end{table}

To evaluate the clustering performance, we compute the agreement between the estimated and true clustering profiles using the Adjusted Rand Index (ARI) and the Jaccard Index, where a value of one indicates perfect concordance. Classification error is also considered as a complementary metric to assess the clustering accuracy. The performance of CAPclust, K-means, Hierarchical, MoEClust, and LASIR over $200$ replications is shown in Table~\ref{table:sim_cluster}. Overall, our method consistently outperforms the competing methods across all clustering metrics. As the number of units increases, for the proposed CAPclust approach, the ARI and Jaccard Index improve significantly and the classification error decreases markedly, while the value of these metrics of the competing methods remains similar. Across all simulation settings, clustering performance based on $\log{(\bgamma^\top \bS_i \bgamma)}$ consistently outperforms clustering based on the vectorized lower–triangular correlations. This result is expected since $\log{(\bgamma^\top \bS_i \bgamma)}$ directly corresponds to the variance component specified by the data-generating model along the projecting direction $\bgamma$. Furthermore, when $\log{(\bgamma^\top \bS_i \bgamma)}$ is used as the input, MoEClust performs slightly better than the proposed method. This difference can be attributed to MoEClust being fitted with the true projection $\bgamma$, whereas the proposed procedure estimates $\bgamma$ from the data.

For the evaluation of choosing the number of clusters, we employ the average BIC criterion discussed in Section \ref{sub:ncluster}. Our proposed method demonstrates a commendable accuracy of $89.5\%$ in correctly identifying \(K=2\) clusters. Additional details of the simulation are provided in Section~\ref{sup:optim_clust} of the supplementary materials.

\begin{sidewaystable}[ht]
\caption{\label{table:sim_cluster}
Clustering performance for $K=2$ clusters over $200$ replications in the simulation study.
ARI: adjusted Rand index; Jaccard: Jaccard index; $\text{Clus}_{\text{error}}$: clustering error.
low.tri: vectorization of the lower-triangular part of the covariance matrix; log: $\log(\gamma \bS_{i}\gamma)$ as input.}
\centering
\scriptsize
\setlength{\tabcolsep}{3pt}
\renewcommand{\arraystretch}{1.05}
\begin{tabular}{c l r r r c r r r c r r r c r r r}
\hline
\multirow{3}{*}{$(p,n,T)$} & \multirow{3}{*}{Method} &
\multicolumn{7}{c}{\textbf{Dim 2}} & &
\multicolumn{7}{c}{\textbf{Dim 2 \& Dim 4 (intercept only)}}\\
\cline{3-9}\cline{11-17}
& & \multicolumn{3}{c}{intercept only} & & \multicolumn{3}{c}{with $\bw$} & &
\multicolumn{3}{c}{Dim 2} & & \multicolumn{3}{c}{Dim 4} \\
\cline{3-5}\cline{7-9}\cline{11-13}\cline{15-17}
& & Jaccard & ARI & $\text{Clus}_{\text{error}}$
  & & Jaccard & ARI & $\text{Clus}_{\text{error}}$
  & & Jaccard & ARI & $\text{Clus}_{\text{error}}$
  & & Jaccard & ARI & $\text{Clus}_{\text{error}}$\\
\hline
\multirow{8}{*}{$(50,50,100)$}
 & CAPclust               & 0.908 & 0.846 & 0.047 & & 0.900 & 0.792 & 0.068 & & 0.911 & 0.858 & 0.045 & & 0.822 & 0.534 & 0.152 \\
 & K-means (low.tri)      & 0.318 & 0.009 & 0.430 & & 0.376 & 0.005 & 0.436 & & 0.332 & 0.029 & 0.405 & & 0.535 & 0.016 & 0.410 \\
 & K-means (log)          & 0.765 & 0.612 & 0.109 & & 0.820 & 0.645 & 0.099 & & 0.763 & 0.609 & 0.111 & & 0.559 & 0.067 & 0.367 \\
 & Hierarchical (low.tri) & 0.294 & 0.027 & 0.410 & & 0.365 & 0.015 & 0.428 & & 0.316 & 0.041 & 0.394 & & 0.583 & 0.037 & 0.374 \\
 & Hierarchical (log)     & 0.730 & 0.562 & 0.127 & & 0.773 & 0.560 & 0.128 & & 0.728 & 0.555 & 0.129 & & 0.568 & 0.063 & 0.365 \\
 & MoE (low.tri)          & 0.301 & 0.014 & 0.422 & & 0.371 & 0.010 & 0.436 & & 0.327 & 0.039 & 0.397 & & 0.560 & 0.027 & 0.391 \\
 & MoE (log)              & 0.952 & 0.925 & 0.019 & & 0.966 & 0.930 & 0.018 & & 0.953 & 0.927 & 0.019 & & 0.921 & 0.748 & 0.065 \\
 & LASIR                  & 0.369 &-0.009 & 0.462 & & 0.374 &-0.005 & 0.454 & & 0.310 &-0.006 & 0.458 & & 0.479 &-0.002 & 0.448 \\
\hline
\multirow{8}{*}{$(50,100,100)$}
 & CAPclust               & 0.962 & 0.936 & 0.018 & & 0.961 & 0.921 & 0.024 & & 0.952 & 0.923 & 0.021 & & 0.830 & 0.591 & 0.131 \\
 & K-means (low.tri)      & 0.335 & 0.003 & 0.454 & & 0.387 & 0.002 & 0.456 & & 0.328 & 0.004 & 0.454 & & 0.462 & 0.007 & 0.445 \\
 & K-means (log)          & 0.799 & 0.652 & 0.096 & & 0.836 & 0.663 & 0.092 & & 0.798 & 0.652 & 0.096 & & 0.543 & 0.090 & 0.035 \\
 & Hierarchical (low.tri) & 0.326 & 0.022 & 0.425  & & 0.408 & 0.021 & 0.429 & & 0.312 & 0.021 & 0.426 & & 0.500 & 0.004 & 0.430 \\
 & Hierarchical (log)     & 0.763 & 0.595 & 0.117 & & 0.791 & 0.582 & 0.120 & & 0.763 & 0.594 & 0.117 & & 0.524 & 0.074 & 0.367 \\
 & MoE (low.tri)          & 0.312 & 0.003 & 0.453  & & 0.387 & 0.002 & 0.457 & & 0.494   & 0.002 & 0.455  & & 0.470 & 0.005  & 0.442  \\
 & MoE (log)              & 0.961 & 0.934 & 0.017 & & 0.969 & 0.936 & 0.016 & & 0.961 & 0.935 & 0.016 & & 0.913 & 0.756 & 0.063 \\
 & LASIR                  & 0.320 &-0.003 & 0.465 & & 0.366 &-0.005 & 0.474 & & 0.320 &-0.003 & 0.465 & & 0.448 & 0.003 & 0.451 \\
\hline
\end{tabular}
\end{sidewaystable}


\section{The ADNI Study}
\label{sec:application}

We apply the proposed approach to the resting-state fMRI data collected in the Alzheimer's Disease Neuroimaging Initiative (ADNI, \url{adni.loni.usc.edu}). The ADNI is a large-scale, multi-center longitudinal study launched in 2004 with the goal of identifying and validating biomarkers for AD and measuring the progression of mild cognitive impairment (MCI) and AD.
The current study analyzes the resting-state fMRI data collected from $n=162$ individuals at the screening visit. Among these, $58$ are cognitive normal individuals ($33$ Female), $73$ diagnosed with MCI ($34$ Female), and $31$ with AD ($16$ Female). To minimize the scanner effect, only data collected from a Philips scanner with $\mathrm{TR}=3$ seconds are analyzed. The fMRI time courses are extracted from $p=75$ brain regions ($60$ cortical and $15$ subcortical regions) using the Harvard-Oxford Atlas in FSL~\citep{smith2004advances}, where these regions are grouped into ten functional modules. This modular information is used to sparsify the loading profile ($\bgamma$) using the fused lasso~\citep{tibshirani2005sparsity} after identifying the components to serve better interpretation. For all individuals, the number of fMRI volumes is $T_{i}=T=134$. These fMRI volumes are standardized to mean zero and unit variance, denoted as $\by_{it}\in\mathbb{R}^{p}$, for $t=1,\dots,T_{i}$ and $i=1,\dots,n$. It is assumed that $\by_{it}$ follows a normal distribution with mean zero and covariance matrix $\bSigma_{i}$. These covariance matrices reveal the architecture of the coherence in brain activity at rest, the so-called resting-state brain functional connectivity. The objective is to cluster individuals based on this brain coherence and identify the associated brain subnetworks. The covariates considered in the parsimonious model ($\bx$) include gender, age, and their interaction, and years of education ($q_{1}=5$). Both age and years of education are centered to mean zero. The same set of covariates is also considered in the mixing proportion model ($\bw$ with $q_{2}=5$).
Imposed model assumptions are examined in Section~\ref{appendix:sec:application} of the supplementary materials.

\begin{figure}
    \begin{center}
        \subfloat[C1]{\includegraphics[width=0.3\textwidth]{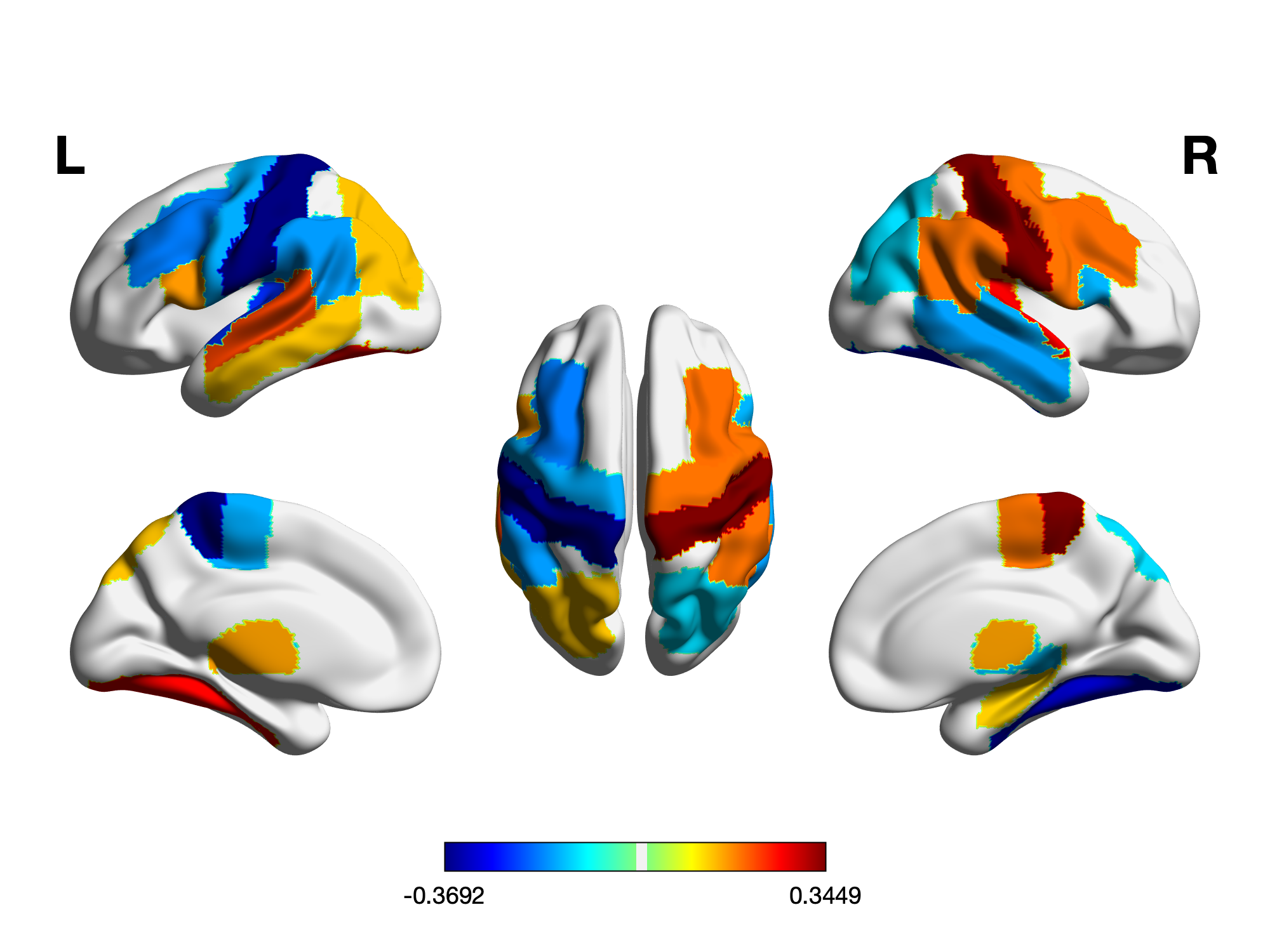}}
        \enskip{}
        \subfloat[C2]{\includegraphics[width=0.3\textwidth]{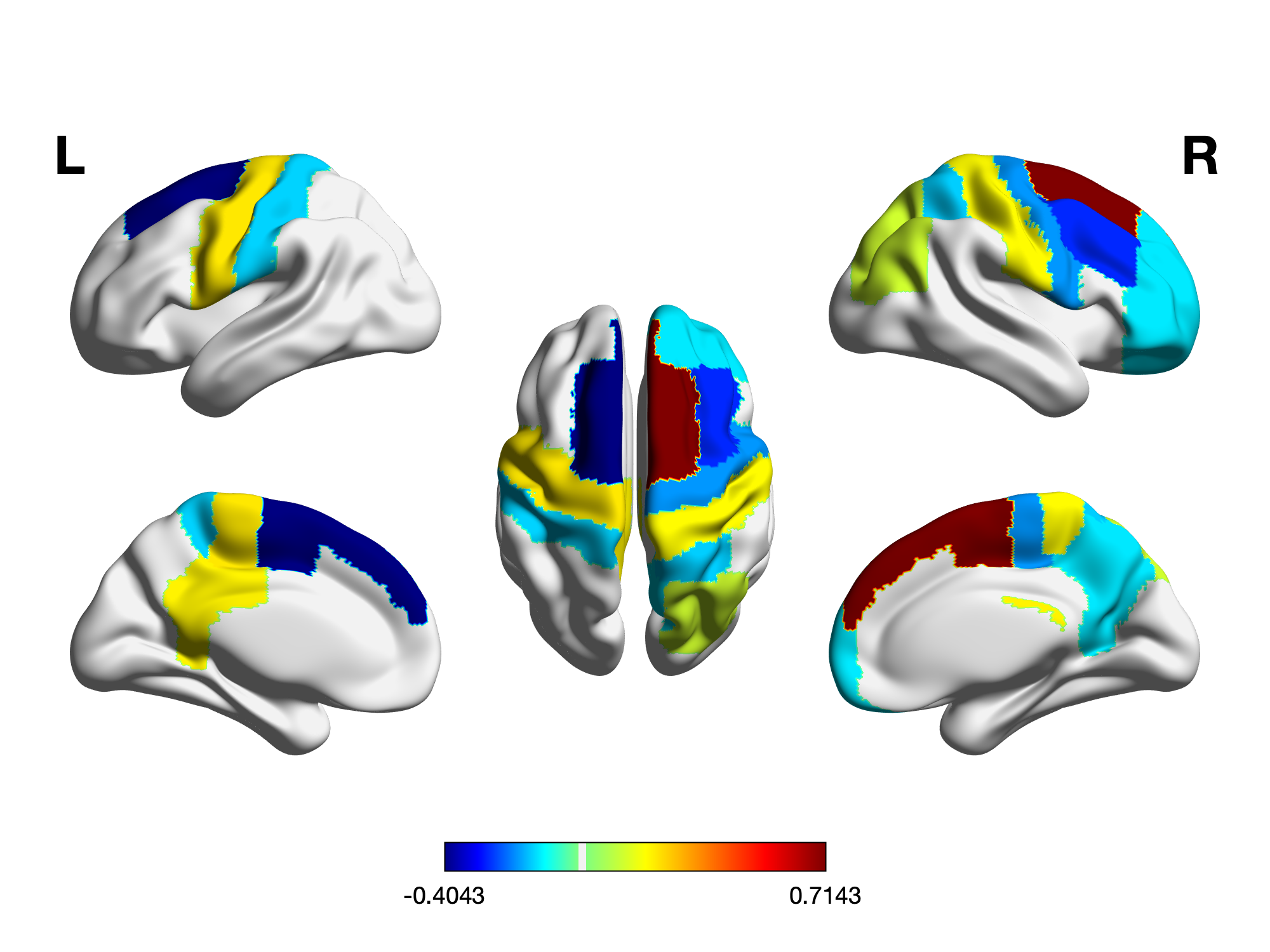}}
        \enskip{}
        \subfloat[C3]{\includegraphics[width=0.3\textwidth]{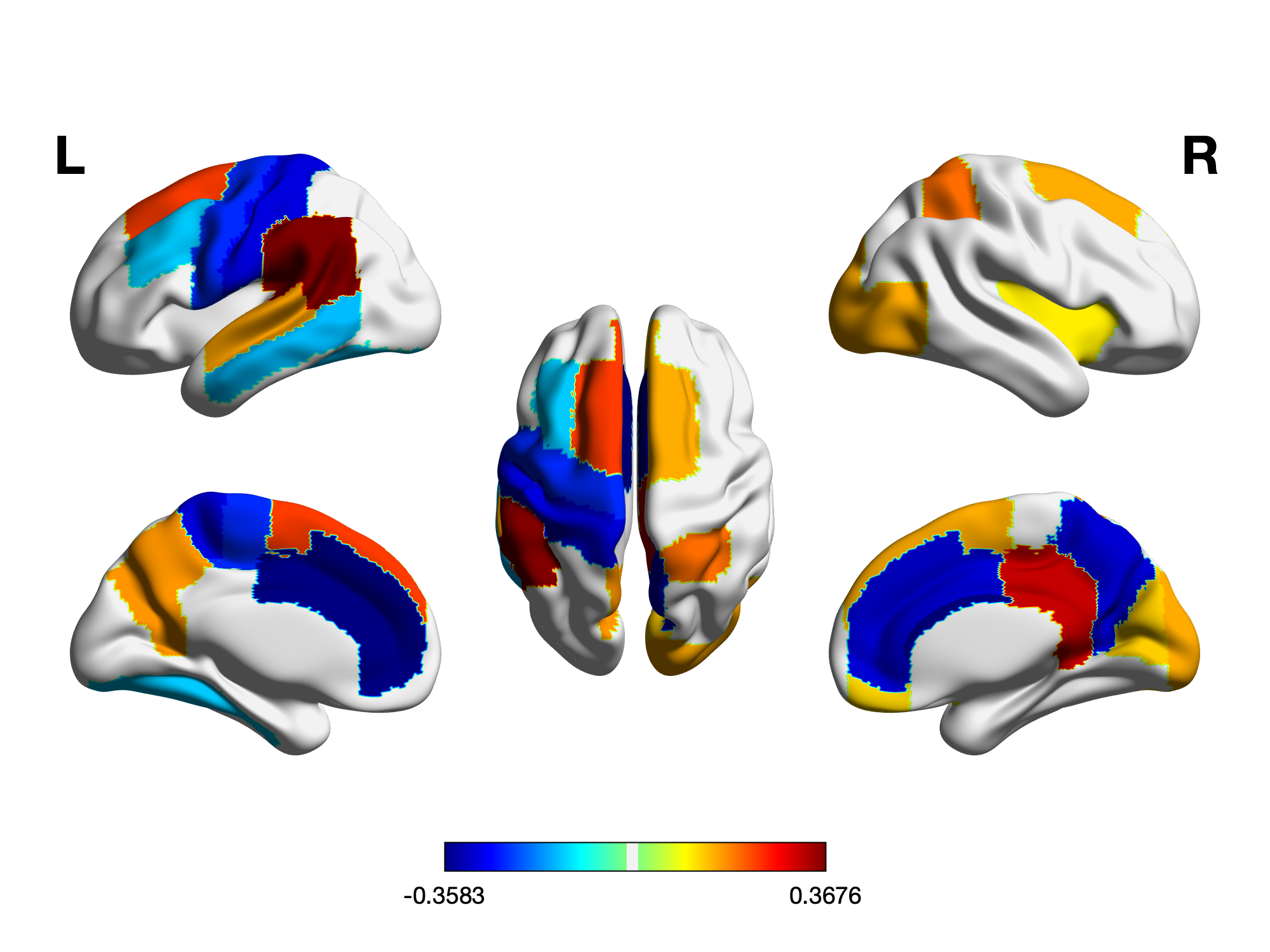}}
    \end{center}
    \caption{\label{fig:adni_brain}The brain map of the three identified components in the ADNI analysis.}
\end{figure}

\begin{figure}
    \begin{center}
        \subfloat[Diagnosis$\times$Age: C1]{\includegraphics[width=0.3\textwidth]{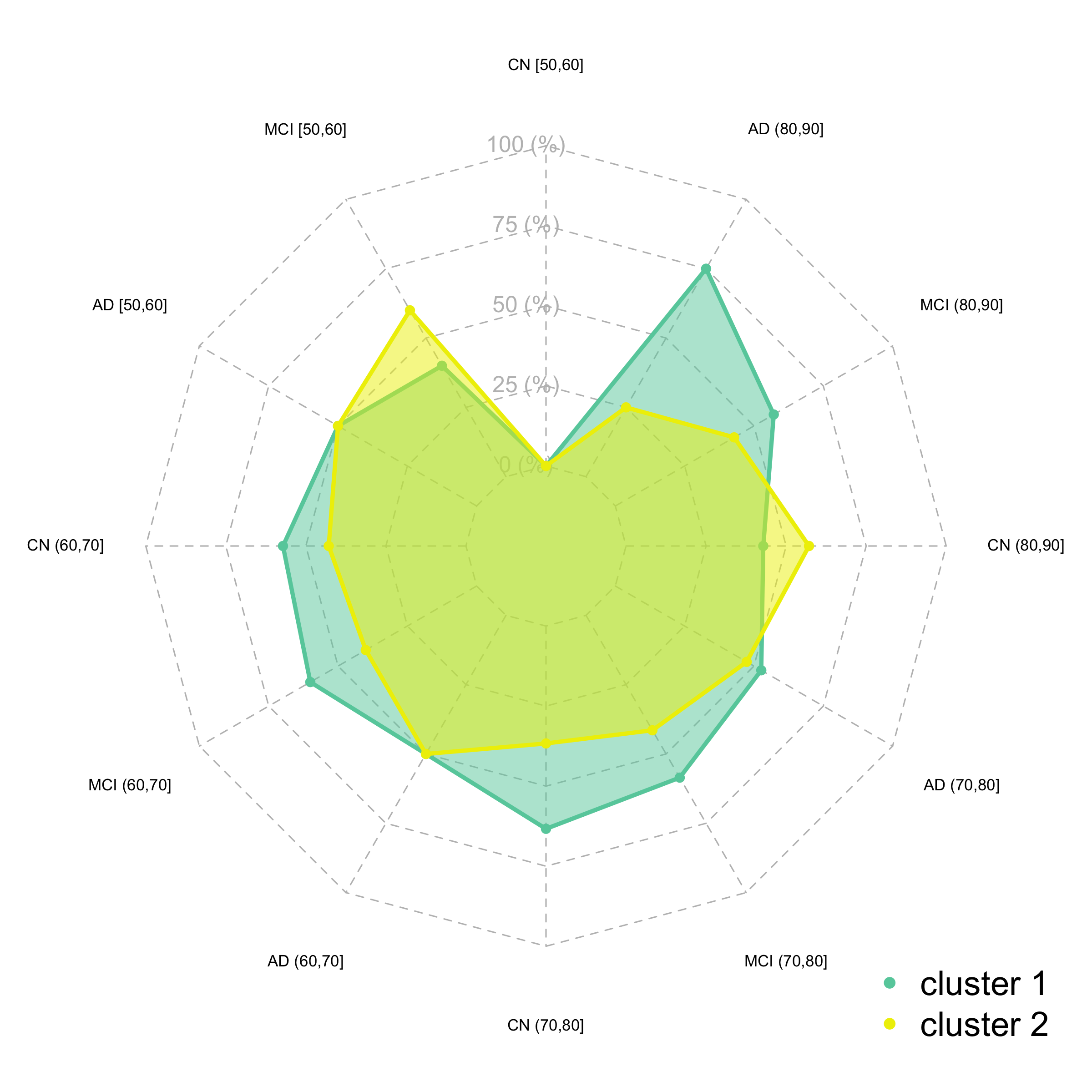}}
        \enskip{}
        \subfloat[Diagnosis$\times$Age: C2]{\includegraphics[width=0.3\textwidth]{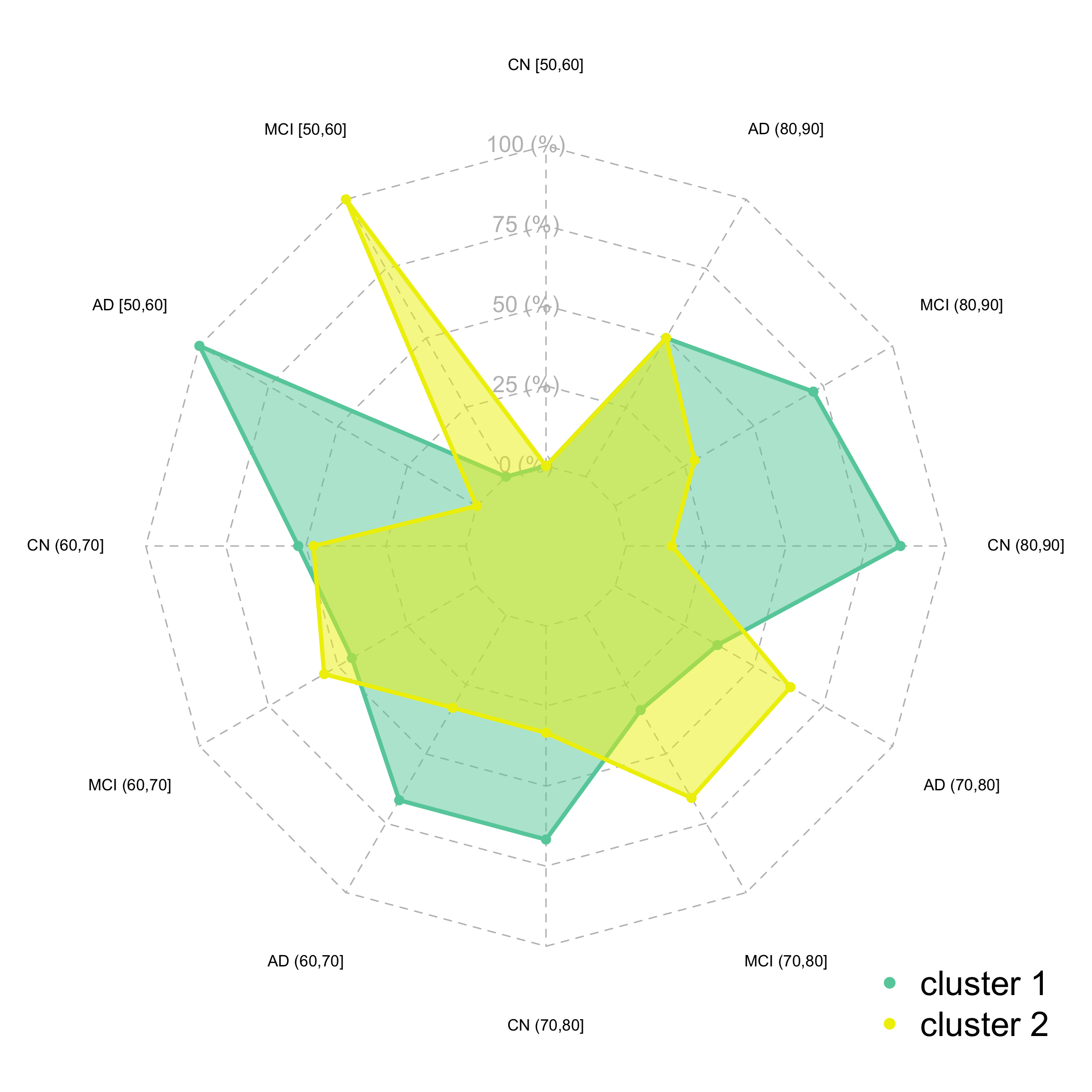}}
        \enskip{}
        \subfloat[Diagnosis$\times$Age: C3]{\includegraphics[width=0.3\textwidth]{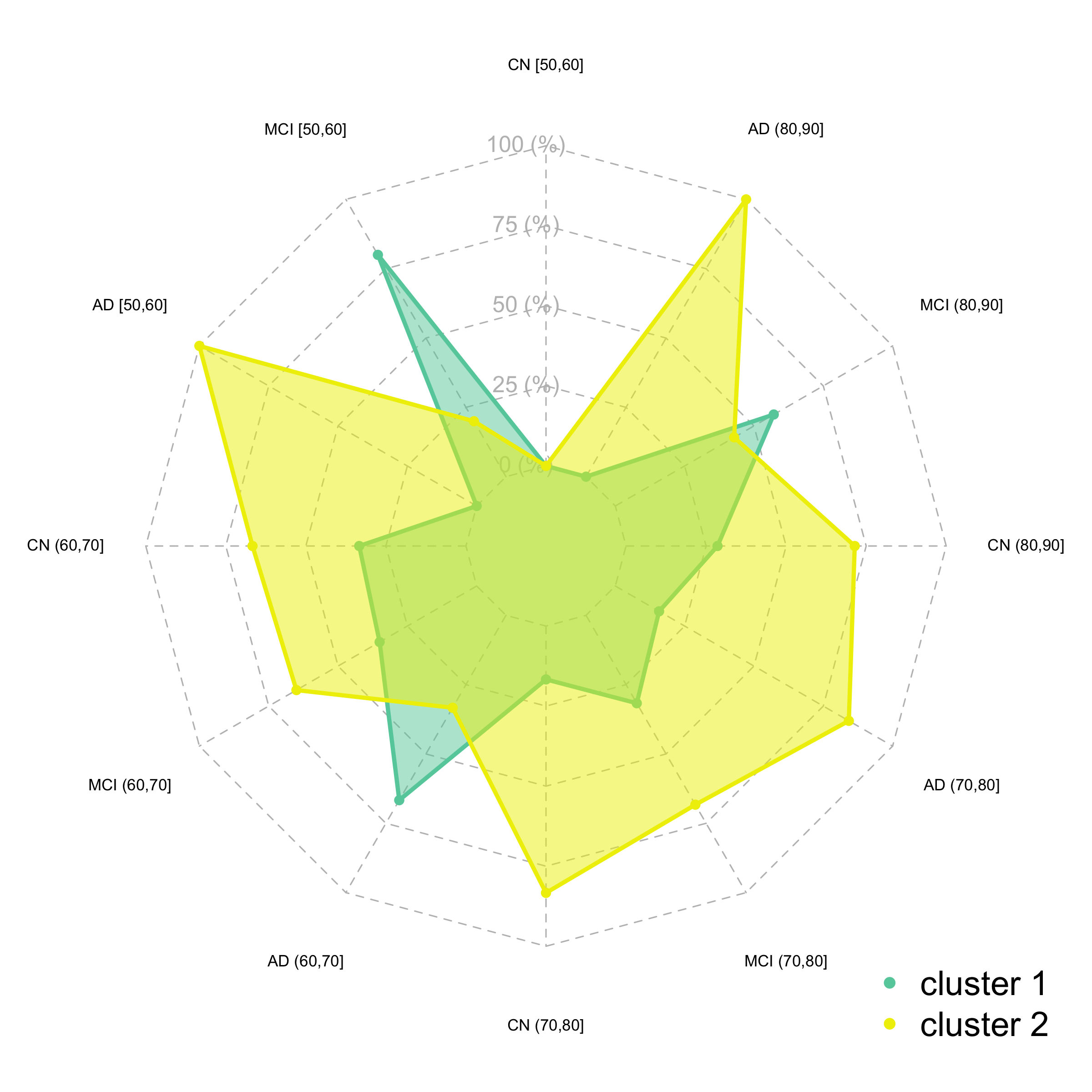}}

        \subfloat[Diagnosis$\times$Gender: C1]{\includegraphics[width=0.3\textwidth]{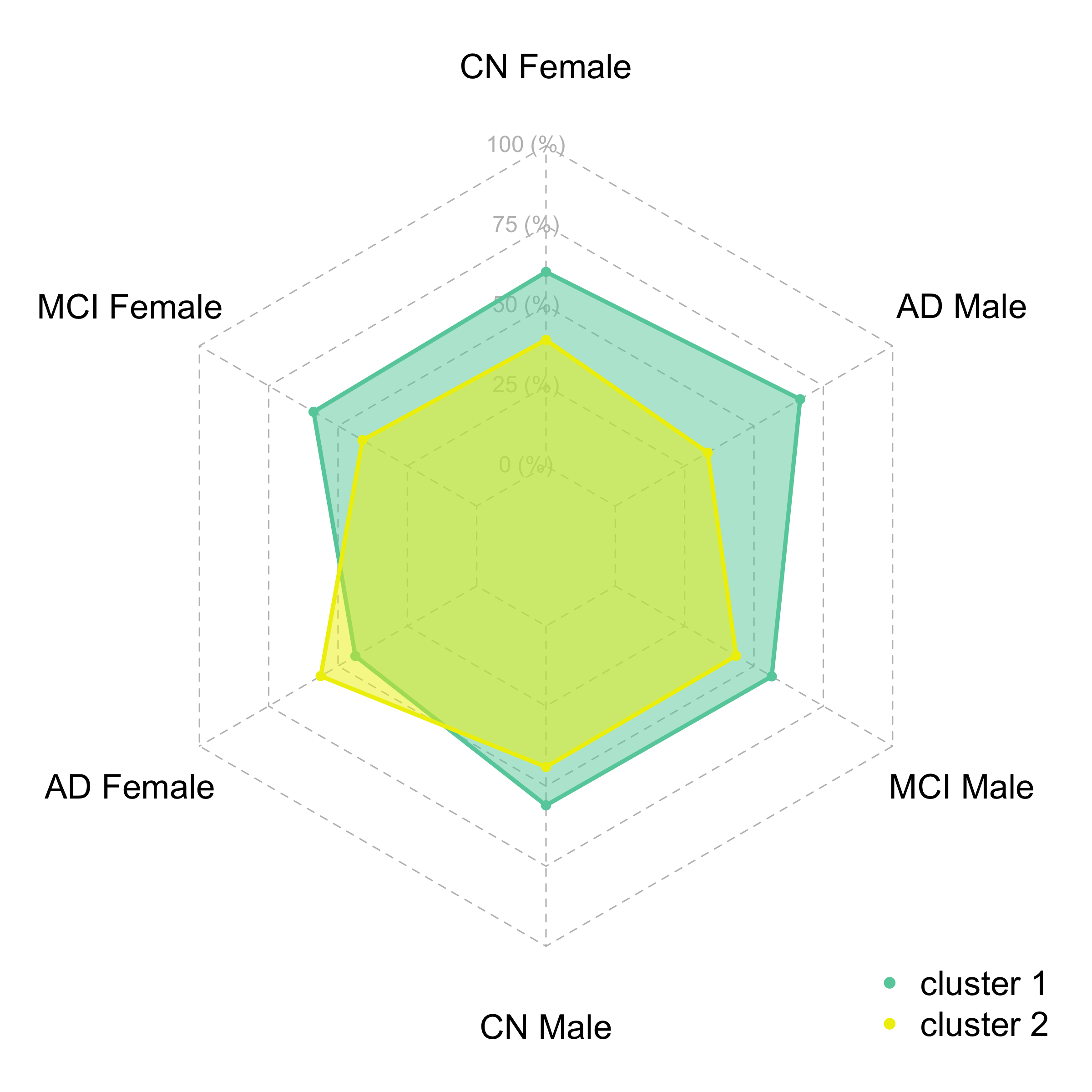}}
        \enskip{}
        \subfloat[Diagnosis$\times$Gender: C2]{\includegraphics[width=0.3\textwidth]{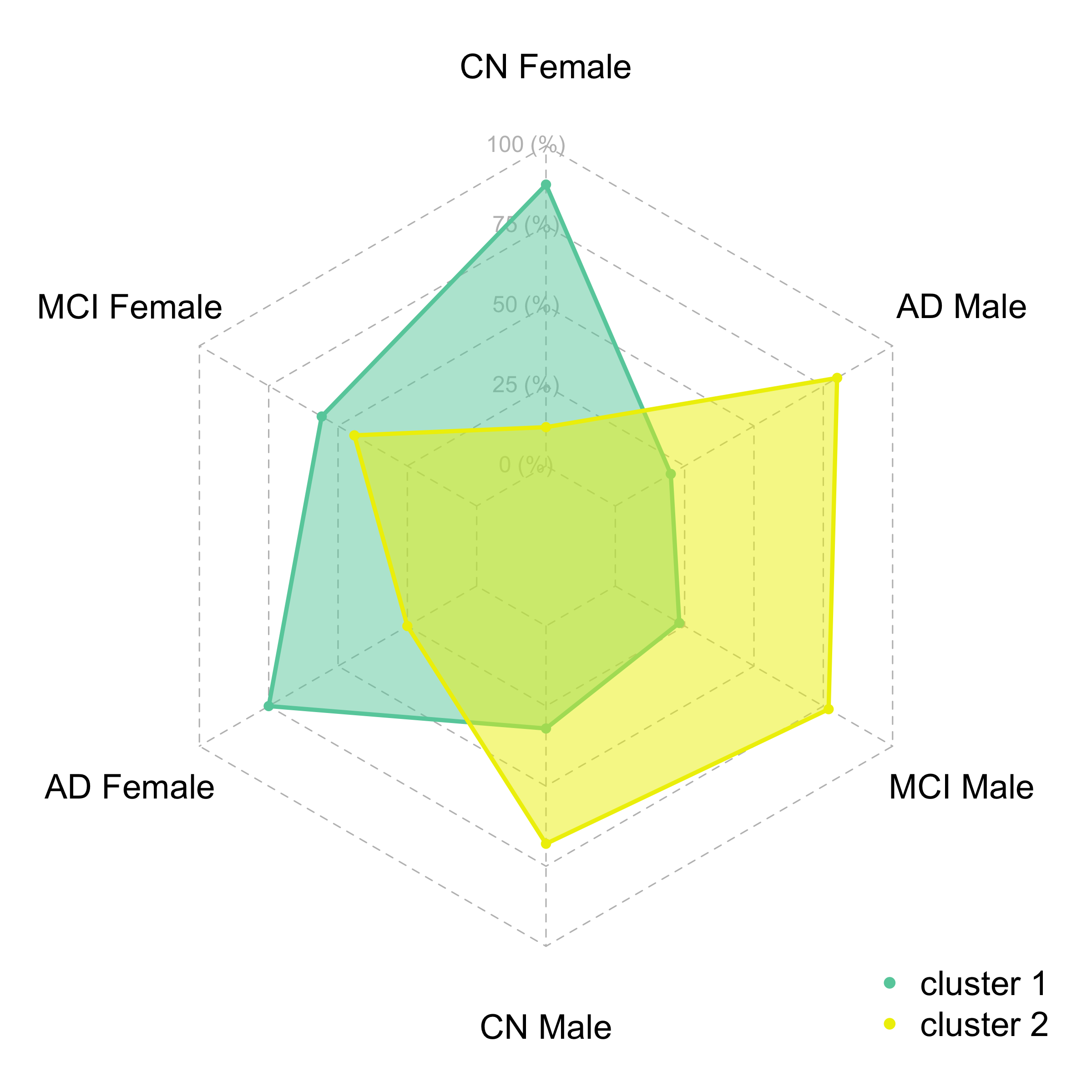}}
        \enskip{}
        \subfloat[Diagnosis$\times$Gender: C3]{\includegraphics[width=0.3\textwidth]{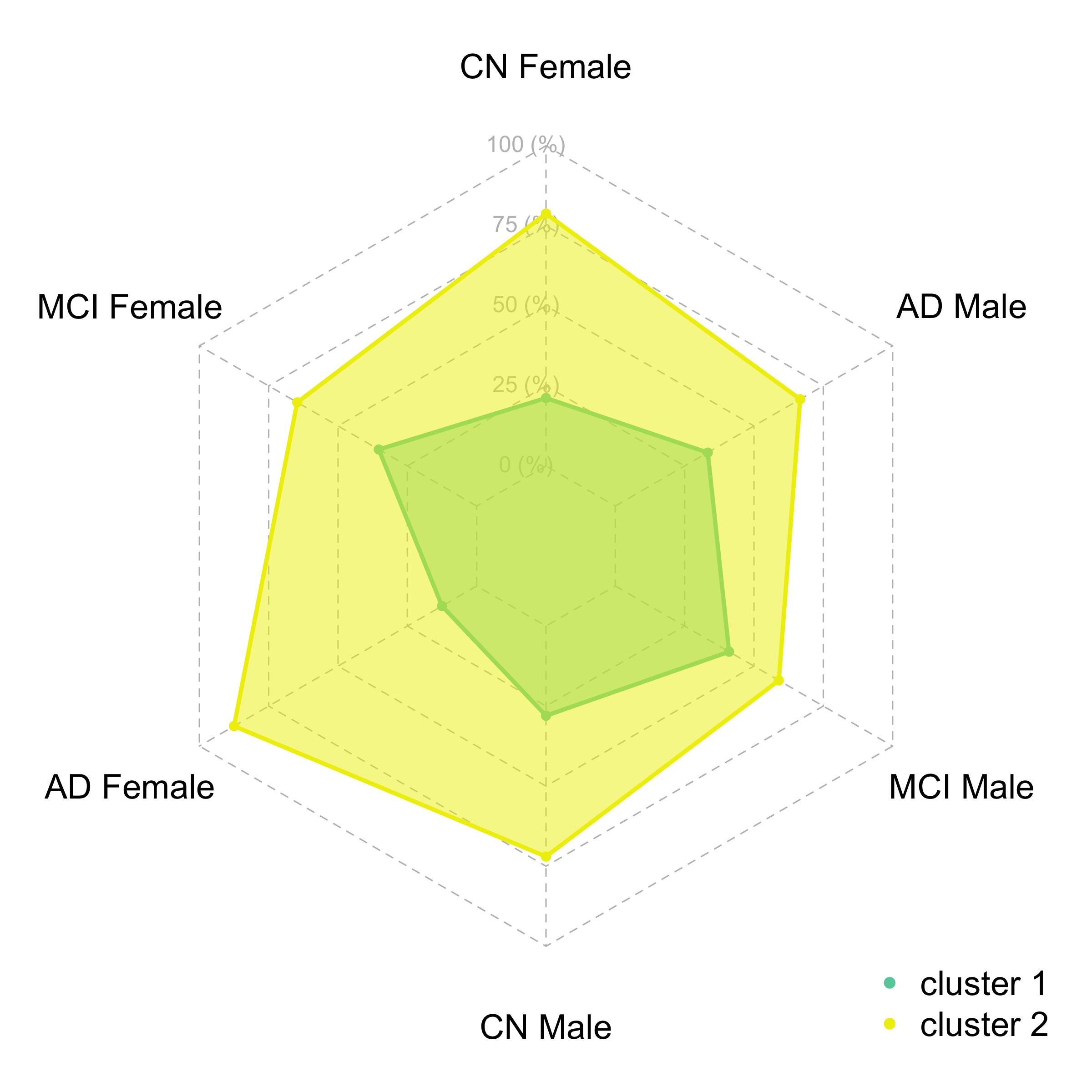}}
    \end{center}
    \caption{\label{fig:adni_cluster}Individual clustering polar plot of the three identified components compared to (a)--(c) Diagnosis$\times$Age and (d)--(f) Diagnosis$\times$Gender. }
\end{figure}

Using the BIC criterion discussed in Section~\ref{sub:ncluster}, $K=2$ clusters are selected. Using the $\mathrm{DfD}$ metric in~\eqref{eq:DfD} and setting the threshold at two, the proposed approach identifies three components. Table~\ref{table:adni} presents the coefficient estimates and contrast comparisons in the variance model~\eqref{eq:model_var}. For C1, the variance is significantly associated with years of education in both cluster 1 and cluster 2. For C2, the variance is significantly positively associated with age in females in both cluster 1 and cluster 2, but only significantly negatively associated with age in males in cluster 2. In cluster 2, there is also a significant gender difference at the average age ($72.7$ years old). The age and gender interaction is significant in both clusters. For C3, in both clusters, the variance is significantly positively associated with age in females and the age$\times$gender interaction is significant.
Figure~\ref{fig:adni_brain} shows the regions in brain maps after sparsifying the loadings. 
C1 covers regions from the temporal lobe and default mode network (DMN), including the pars opercularis, temporal gyrus, middle temporal gyrus, fusiform, and supramarginal gyrus, with the function of language processing and highly related to education~\citep{panda2014unraveling,perry2017independent}.
C2 is a component that mainly consists of regions from the DMN and C3 is a component that consists of regions from the DMN, somato-motor, dorsal, and ventral attention networks. Gender differences have been observed in these networks among aging populations~\citep{cavedo2018sex,li2021sex}.
Figure~\ref{fig:adni_cluster} presents the clustering polar plot when comparing to either diagnosis$\times$age or diagnosis$\times$gender. Here, for easy presentation, we break age into categories. As C1 is associated with years of education only, no separation between the two clusters is observed when considering diagnosis, age, and gender. For C2 and C3, a separation of the two clusters is observed.

\begin{sidewaystable}
    \caption{\label{table:adni}Coefficients estimate and contrast comparisons in the variance model ($\bbeta$) of the three identified components in the ADNI analysis. The numbers in the parentheses are the $95\%$ confidence interval from $500$ bootstrap samples.}
    \begin{center}
        \resizebox{\textwidth}{!}{
        \begin{tabular}{l r r c r r c r r}
            \hline
            & \multicolumn{2}{c}{C1} && \multicolumn{2}{c}{C2} && \multicolumn{2}{c}{C3} \\
            \cline{2-3}\cline{5-6}\cline{8-9}
            & \multicolumn{1}{c}{Cluster 1} & \multicolumn{1}{c}{Cluster 2} && \multicolumn{1}{c}{Cluster 1} & \multicolumn{1}{c}{Cluster 2} && \multicolumn{1}{c}{Cluster 1} & \multicolumn{1}{c}{Cluster 2} \\
            \hline
            Age (Female) & $0.056$ ($-0.100$, $0.212$) & $0.050$ ($-0.147$, $0.247$) && $0.105$ ($0.042$, $0.167$) & $0.202$ ($0.036$, $0.368$) && $0.352$ ($0.215$, $0.489$) & $0.163$ ($0.104$, $0.222$)\\
            Age (Male) & $0.203$ ($-0.011$, $0.418$) & $-0.063$ ($-0.243$, $0.117$) && $-0.233$ ($-0.496$, $0.031$) & $-0.101$ ($-0.176$, $-0.025$) && $-0.042$ ($-0.137$, $ 0.053$) & $-0.046$ ($-0.116$, $0.024$)\\
            Male$-$Female & $-0.093$ ($-0.372$, $0.186$) & $-0.184$ ($-0.550$, $0.182$) && $1.256$ ($0.835$, $1.677$) & $-1.298$ ($-1.535$, $-1.062$) && $-0.156$ ($-0.408$, $ 0.094$) & $-0.063$ ($-0.160$, $0.033$)\\
            Gender$\times$Age & $0.148$ ($-0.372$, $0.186$) & $-0.113$ ($-0.377$, $0.150$) && $-0.337$ ($-0.610$, $-0.065$) & $-0.302$ ($-0.493$, $-0.112$) && $-0.340$ ($-0.554$, $ -0.234$) & $-0.209$ ($-0.298$, $-0.120$)\\
            Education & $0.437$ ($0.260$, $0.613$) & $-0.871$ ($-1.058$, $-0.684$) && $-0.086$ ($-0.167$, $-0.006$) & $0.071$ ($-0.028$, $0.169$)&& $-0.029$ ($-0.124$, $0.066$) & $-0.041$ ($-0.091$, $0.008$)\\
            \hline
        \end{tabular}
        }
    \end{center}
\end{sidewaystable}


\section{Discussion}
\label{sec:discussion}
We propose a parsimonious clustering model that integrates a MoE and covariance regression framework to cluster individuals based on covariance matrices. The model assumes a common linear projection structure across covariance matrices, allowing for a flexible yet interpretable clustering solution. A generalized linear model with a logarithmic link function is applied to model the variance while accounting for covariates. By clustering over data in the projected space, the proposed framework allows for different clustering patterns across different projection components of the covariance matrices.
To evaluate the performance of the proposed approach, we conduct simulation studies to assess its robustness, ability to select the correct number of clusters, and clustering accuracy compared to existing methods. We apply the framework to resting-state fMRI data from the ADNI study to identify subgroups based on brain coherence. Our approach effectively uncovers meaningful clusters, identifying distinct brain subnetworks associated with cognitive function and demographic factors. 

There are several directions for future research of the current proposal. First, our current approach assumes a common diagonalization for the covariance matrices, which may be overly restrictive when applied to fMRI data. Simulation studies suggest the robustness of the proposed approach to partial common diagonalization. However, the theoretical study of estimation consistency under this relaxation is challenging. We leave it to one of our future directions.
Second, the current analysis does not address inference on the projection vectors due to the theoretical and computational challenges. Future work should focus on addressing these challenges to improve the inference procedures. Lastly, our proposed method can be extended to high-dimensional settings, on both the dimension of the covariance matrices and the covariates. Incorporating regularization techniques in such cases would provide a valuable enhancement to the existing framework.
Though the proposed approach enables different clustering patterns across different projecting components, the number of clusters is assumed to be identical. Another direction of future research is to allow the number of clusters to vary across components.


\clearpage

\appendix
\counterwithin{figure}{section}
\counterwithin{table}{section}
\counterwithin{equation}{section}
\counterwithin{lemma}{section}
\counterwithin{theorem}{section}

\section*{Appendix}

This Appendix collects the technical proof of the theorems in the main text, additional technical details, and additional numerical results.


\section{A solution for solving~\eqref{eq:solve_gamma}}
\label{appendix:sec:solve_gamma}

To solve for $\bgamma$ in~\eqref{eq:solve_gamma}, it is equivalent to find the eigenvectors and eigenvalues of $\bA$ with respect to $\bH$. We assume there is a solution eigenvector $\bgamma_{0}$ with unit norm, $\|\bgamma_{0}\|_{2}=1$. Since $\bH$ is positive definite, let $\bgamma=\bH^{-1 / 2} \bgamma_{0}$, then
\[
  \bgamma^{\top} \bH \bgamma=\bgamma_{0}^{\top} \bH^{-1 / 2} \bH \bH^{-1 / 2} \bgamma_{0}=\bgamma_{0}^{\top} \bgamma_{0}=1,
\]
which satisfies the constraint condition. Substitute $\bgamma$ with $\bgamma=\bH^{-1 / 2} \bgamma_{0}$ in~\eqref{eq:solve_gamma}, we obtain
\[
\bA\bH^{-1 / 2} \bgamma_{0}-\lambda \bH\bH^{-1 / 2} \bgamma_0=\boldsymbol{\mathrm{0}}.
\]
\[
\Rightarrow \quad \bH^{-1 / 2} \bA\bH^{-1 / 2} \bgamma_{0}=\lambda \bgamma_{0}.
\]
Thus, $\bgamma_{0}$ is the eigenvector of matrix $\bH^{-1 / 2} \bA \bH^{-1 / 2}$ with $\lambda$ as the corresponding eigenvalue. Therefore, the update for $\bgamma$ is given by $\bgamma=\bH^{-1/2}\bgamma_{0}$.


\section{Theory and Proof}
\subsection{Regularity Conditions}
\label{appendix:sec:proof}

Let \( f(\bv; \tilde{\bTheta}) \) be the joint density function of \( \bV = (\bx, \bz, \bw)\), with the parameter space \( \tilde{\bTheta} \in \bOmega \). 
The conditional density function of $\bz$ given $\bx$ follows the MoE model~\eqref{eq:model_mix}. In the following regularity conditions, we denote the parameter vector as \(  \tilde{\bTheta} = (\psi_1, \psi_2, \dots, \psi_{v}) \), where \( v\) represents the total number of parameters in the model. Similar conditions were also imposed in \citet{khalili2010new} for MoE models. 

\begin{description}
    \item \textbf{R1:} The density \( f(\bv; \tilde{\bTheta}) \) has common support in \( \bv \) for all \( \tilde{\bTheta} \in \bOmega \), and \( f(\bv; \tilde{\bTheta}) \) is identifiable with respect to \( \tilde{\bTheta} \).
    
    \item \textbf{R2:} There exists an open subset \( \bOmega^* \subset \bOmega \) containing the true parameter \( \tilde{\bTheta}^* \) such that for almost all \( \bv \), \( f(\bv; \tilde{\bTheta}) \) admits third-order partial derivatives with respect to \( \tilde{\bTheta} \in \bOmega^* \).
    
    \item \textbf{R3:} For all \( j, l = 1, 2, \ldots, v \), the following conditions hold:
    \[
    \mathbb{E}_0 \left[ \frac{\partial}{\partial \psi_j} \log f(\bV; \tilde{\bTheta}) \right] = 0,
    \]
    \[
    \mathbb{E}_0 \left[ \frac{\partial}{\partial \psi_j} \log f(\bV; \tilde{\bTheta}) \frac{\partial}{\partial \psi_l} \log f(\bV; \tilde{\bTheta}) \right] = \mathbb{E}_0 \left[ -\frac{\partial^2}{\partial \psi_j \partial \psi_l} \log f(\bV; \tilde{\bTheta}) \right].
    \]
    
    \item \textbf{R4:} The Fisher information matrix \( \mathcal{I}_n(\tilde{\bTheta}) \) is finite and positive definite at \( \tilde{\bTheta} = \tilde{\bTheta}^* \):
    \[
    \mathcal{I}_n(\tilde{\bTheta}) = \mathbb{E} \left\{ \left[ \frac{\partial}{\partial \tilde{\bTheta}} \log f(\bV; \tilde{\bTheta}) \right] \left[ \frac{\partial}{\partial \tilde{\bTheta}} \log f(\bV; \tilde{\bTheta}) \right]^\top \right\}.
    \]
    
    \item \textbf{R5:} There exists an integrable function \( \mathcal{B}(\bv) \), such that for all \( \tilde{\bTheta} \in \bOmega \) and for any indices \( j, l, m \):
    \[
    \left| \frac{\partial f(\bv; \tilde{\bTheta})}{\partial \psi_j} \right| \leq \mathcal{B}(\bv), \quad \left| \frac{\partial^2 f(\bv; \tilde{\bTheta})}{\partial \psi_j \partial \psi_l} \right| \leq \mathcal{B}(\bv), \quad \left| \frac{\partial^3 \log f(\bv; \tilde{\bTheta})}{\partial \psi_j \partial \psi_l \partial \psi_m} \right| \leq \mathcal{B}(\bv).
    \]
    
 \end{description}   

\subsection{Proof of Theorem~\ref{thm:asymp_gammaknown}}
\label{sup:asymp_gammaknown}

Define $M(\tilde{\bTheta}) = \mathbb{E}_0\left\{ \tilde{\ell}_i(\tilde{\bTheta}) \right\}$, the expected log-pseudo-likelihood under the true parameter $\tilde{\bTheta}^*$. By regularity condition~R1, $M(\tilde{\bTheta})$ has a unique maximizer at $\tilde{\bTheta}^*$. Under regularity conditions~R2--R5, the function $\tilde{\ell}_i(\tilde{\bTheta})$ is continuous and dominated by an integrable envelope. 
By the uniform law of large numbers,
$$
\sup_{\tilde{\bTheta} \in \Omega^*} \left| \tilde{\mathcal L}_n(\tilde{\bTheta}) - M(\tilde{\bTheta}) \right| \xrightarrow{\mathcal{P}} 0.
$$
where $\tilde{\ell}_n=\sum_{i=1}^n \tilde{\ell}_i$ is the sample log pseduo-likelihood, $\tilde{\mathcal L}_n=\tilde{\ell}_n/n$ is the average log pseduo-likelihood, and $\overset{\mathcal{P}}{\longrightarrow}$ means converge in probability.
Then, by the argmax theorem, any maximizer $\hat{\tilde{\bTheta}}$ of $\tilde{\ell}_n$ satisfies $\hat{\tilde{\bTheta}} \xrightarrow{\mathcal{P}} \tilde{\bTheta}^*$.

To prove the root-$n$ convergence rate, a Taylor expansion of the score around $\tilde{\bTheta}^*$ yields
$$
0=S_n\left(\tilde{\bTheta}^*\right)+H_n(\bar{\bTheta})\left(\hat{\tilde{\bTheta}}-\tilde{\bTheta}^*\right)
$$
where $S_n\left(\tilde{\bTheta}^*\right)$ and $H_n(\bar{\bTheta})$ denote the first and second order derivative with respect to $\tilde{\bTheta}$. By simple algebra,
$$
\sqrt{n}\left(\hat{\tilde{\bTheta}}-\tilde{\bTheta}^*\right)=\left[-\frac{1}{n} H_n(\bar{\bTheta})\right]^{-1} \frac{1}{\sqrt{n}} S_n\left(\tilde{\bTheta}^*\right).
$$
By R3 and R4, $\mathbb{E}_0\left\{S_i\left(\tilde{\bTheta}^*\right)\right\}=0$ and $\mathcal{I}(\tilde{\bTheta}^*)$ is positive definite, then by a multivariate central limit theorem under R2 and R5,
\[
    \frac{1}{\sqrt{n}} S_n\left(\tilde{\Theta}^*\right) \overset{\mathcal{D}}{\longrightarrow} \mathcal{N}\left(0, \mathcal{I}\left(\tilde{\Theta}^*\right)\right).
\]
Moreover, by a law of large numbers and the consistency established above,
\[
-\tfrac{1}{n}H_n(\bar{\boldsymbol{\Theta}})\ \xrightarrow{p}\ \mathcal I(\tilde{\boldsymbol{\Theta}}^*).
\]
By Slutsky’s theorem,
\[
\sqrt{n}\big(\hat{\tilde{\boldsymbol{\Theta}}}-\tilde{\boldsymbol{\Theta}}^*\big)
\overset{\mathcal{D}}{\longrightarrow}
\mathcal N\!\big(\boldsymbol{0},\,\mathcal I(\tilde{\boldsymbol{\Theta}}^*)^{-1}\big).
\]
Hence the RHS is $O_p(1)$, and it proves $\left\|\hat{\tilde{\bTheta}}-\tilde{\bTheta}^*\right\|=O_p\left(n^{-1 / 2}\right)$.

\subsection{Proof of Theorem~\ref{thm:asymp_gammaunknown}}
\label{sup:asymp_gammaunknown}

In addition to the regularity conditions R1--R5, Assumptions A1--A3 ensure that the model is identifiable in $\bgamma$ up to a sign flip. Define $Q(\bTheta)= E_0\{\ell_i(\bTheta)\} = E_0\{\log f(V;\bTheta)\}$,
the expected log-likelihood per subject. By Condition R1 and Assumptions A2--A3, $Q(\bTheta)$ attains a unique global maximum at the true parameter $\bTheta^*=(\bgamma^*,\tilde{\bTheta}^*)$. Moreover, under Conditions R2--R5 and Assumption A1, the sample criterion $\mathcal{L}_n(\bTheta)=\ell_n(\bTheta)/n$ converges uniformly in probability to $Q(\bTheta)$ as $n\to\infty$. By the argmax theorem, any maximizer $\hat{\bTheta}=(\hat{\bgamma},\hat{\tilde{\bTheta}})$ is consistent: $\hat{\bTheta}\xrightarrow{\mathcal{P}}\bTheta^*$, and thus $\hat{\bgamma}\xrightarrow{\mathcal{P}}\bgamma^*$ and $\hat{\tilde{\bTheta}}\xrightarrow{\mathcal{P}}\tilde{\bTheta}^*$. Therefore, the consistency of the estimator with $\bgamma$ is established under the imposed assumptions.

\section{Additional Simulation Study}

\subsection{Choosing the Number of Clusters}
\label{sup:optim_clust}
This section reports the selection of the number of clusters, $K$, using BIC. In the simulation, the true number of clusters for each projecting component is set to be two. With $p=50$ and $n=T=100$, the frequency of identifying the correct number of clusters is 179 out of 200 replications based on lowest average BIC value, indicating that the proposed method recovers the true number of clusters in the majority of simulation runs. Meanwhile, the frequency of identifying the correct projecting components is high, further supporting the reliability of the proposed approach.

\subsection{Partial common eigenstructure}
\label{sup:partial_common}

This section assesses robustness to the violation of the common eigenstructure assumption (Assumption~A2) by modifying the setting of Simulation~(i) in Section~\ref{sec:sim} to partial common eigenstructure. With $p = 50$, the first three eigenvectors are set to be common across subjects, whereas the remaining eigenvectors are generated to be random and subject-specific. Under this setting, component D2 satisfies the model assumption, while D4 does not. The results in Table~\ref{table:sim_est_1} show that the proposed method correctly identifies D2 but does not recover D4 as expected. These findings suggest that our method is robust to the relaxation of complete common eigenstructure to partial common eigenstructure.

\begin{table}
  \caption{\label{table:sim_est_1}
  Performance of estimating the projection ($\bgamma$) and corresponding coefficient ($\bbeta$) over $200$ replications with $p=50$ in the simulation study of partial common eigenstructure. SE: standard error; MSE: mean squared error.}
  \begin{center}
   \resizebox{\textwidth}{!}{
    \begin{tabular}{c c l c rr c rr c rr c rr}
        \hline
        & & \multicolumn{1}{c}{$\hat{\bgamma}$} & \multicolumn{6}{c}{$\hat{\bbeta}_{1}$} & \multicolumn{6}{c}{$\hat{\bbeta}_{2}$} \\
        \cline{3-3}\cline{5-9}\cline{11-15}
        \multicolumn{1}{c}{$(n,T)$} & & & & \multicolumn{2}{c}{Cluster 1} & & \multicolumn{2}{c}{Cluster 2} & & \multicolumn{2}{c}{Cluster 1} & & \multicolumn{2}{c}{Cluster 2} \\
        \cline{5-6}\cline{8-9}\cline{11-12}\cline{14-15}
        & & \multicolumn{1}{c}{\multirow{-2}{*}{$|\langle\hat{\bgamma},\bpi_{j}\rangle|$ (SE)}}  & &
        \multicolumn{1}{c}{Bias} & \multicolumn{1}{c}{MSE} & &
        \multicolumn{1}{c}{Bias} & \multicolumn{1}{c}{MSE} & &
        \multicolumn{1}{c}{Bias} & \multicolumn{1}{c}{MSE} & &
        \multicolumn{1}{c}{Bias} & \multicolumn{1}{c}{MSE} \\
        
            \hline
        \multirow{1}{*}{$(50,100)$} & D2 & $0.986$ ($0.009$) & &
        $-0.057$ & $0.281$ & & $-0.459$ & $1.948$ & &
        $0.015$ & $0.263$ & & $0.090$ & $0.540$ \\
        \cline{3-15}
        \multirow{1}{*}{$(100,100)$} & D2 & $0.997$ ($0.001$) & &
        $-0.013$ & $0.010$ & & $-0.102$ & $0.498$ & &
        $0.004$ & $0.001$ & & $0.026$ & $0.054$ \\
        \hline
    \end{tabular}
    }
  \end{center}
\end{table}

\subsection{Non-Gaussian distributed data}
\label{sup:non_guassian}

This section examines the performance of the proposed method when the data is generated from a non-Gaussian distribution. Specifically, we consider a multivariate $t$-distribution with mean zero, covariance matrix $\bSigma_i$, and degrees of freedom five. The covariance matrices are generated in the same manner as in the Simulation~(i) in Section~\ref{sec:sim}.

Table~\ref{table:sim_est_2} presents the performance of estimating the projection ($\bgamma$) and corresponding coefficient for variance ($\bbeta$). Compared to the results in Table~\ref{table:sim_est}, the estimation bias increases under the multivariate $ t$-distribution setting but is still reasonably well. The performance improves as the sample size increases. These findings suggest that the proposed method is robust to non-Gaussian distributed data to a certain extent.

\begin{table}
  \caption{\label{table:sim_est_2}
  Performance of estimating the projection ($\bgamma$) and corresponding coefficient ($\bbeta$) over $200$ replications with $p=50$ in the simulation study of non-Gaussian distributed data (multivariate $t$ with degrees of freedom of five). SE: standard error; MSE: mean squared error.}
  \begin{center}
   \resizebox{\textwidth}{!}{
    \begin{tabular}{c c l c rr c rr c rr c rr}
        \hline
        & & \multicolumn{1}{c}{$\hat{\bgamma}$} & \multicolumn{6}{c}{$\hat{\bbeta}_{1}$} & \multicolumn{6}{c}{$\hat{\bbeta}_{2}$} \\
        \cline{3-3}\cline{5-9}\cline{11-15}
        \multicolumn{1}{c}{$(n,T)$} & & & & \multicolumn{2}{c}{Cluster 1} & & \multicolumn{2}{c}{Cluster 2} & & \multicolumn{2}{c}{Cluster 1} & & \multicolumn{2}{c}{Cluster 2} \\
        \cline{5-6}\cline{8-9}\cline{11-12}\cline{14-15}
        & & \multicolumn{1}{c}{\multirow{-2}{*}{$|\langle\hat{\bgamma},\bpi_{j}\rangle|$ (SE)}}  & &
        \multicolumn{1}{c}{Bias} & \multicolumn{1}{c}{MSE} & &
        \multicolumn{1}{c}{Bias} & \multicolumn{1}{c}{MSE} & &
        \multicolumn{1}{c}{Bias} & \multicolumn{1}{c}{MSE} & &
        \multicolumn{1}{c}{Bias} & \multicolumn{1}{c}{MSE} \\
        
            \hline
        \multirow{2}{*}{$(50,100)$} & D2 & $0.976$ ($0.015$) & &
        $-0.025$ & $0.051$ & & $-0.278$ & $3.169$ & &
        $-0.005$ & $0.073$ & & $-0.185$ & $12.00$ \\
        & D4 & $0.693$ ($0.057$) & &
        $0.103$ & $0.015$ & & $0.105$ & $0.220$ & &
        $0.092$ & $0.040$ & & $-0.321$ & $0.234$ \\
        \cline{3-15}
        \multirow{2}{*}{$(100,100)$} & D2 & $0.987$ ($0.004$) & &
        $-0.011$ & $0.009$ & & $-0.113$ & $0.392$ & &
        $0.004$ & $0.001$ & & $0.019$ & $0.001$ \\
        & D4 & $0.828$ ($0.089$)  & &
        $-0.167$ & $0.071$ & & $0.131$ & $0.050$ & &
        $0.132$ & $0.051$ & & $-0.172$ & $0.083$ \\
        \hline
    \end{tabular}
    }
  \end{center}
\end{table}

\section{Additional Results for the ADNI Study}
\label{appendix:sec:application}

\subsection{Model assumptions}

This section examines the assumptions introduced in Section~\ref{sub:asymp}. For Assumption~A2 (common eigenstructure), we assess the validity of the assumption empirically as follows. We first compute the average sample covariance matrix over all units and obtain its eigenvectors. We then compare the eigenvectors of the average sample covariance matrix with those of each subject's sample covariance matrix. It is considered to be highly similar for two eigenvectors when their correlation is above $0.5$, allowing for sampling variability and estimation bias. By this criterion, approximately $12\%$ of the eigenvectors show high similarity across subjects, indicating that the common eigenstructure holds only partially. As discussed in Section~\ref{sup:partial_common}, partial common diagonalization does not compromise identification of the common components.

We also assess the Gaussianity assumption for the data. Figure~\ref{table:sim_est_2} shows a normal Q–Q plot and a histogram for the fMRI signal from one brain region of one subject. From the Figure, the marginal distribution appears to be close to normal. We can say that overall, the Gaussian distribution assumption is satisfied.

\begin{figure}
    \begin{center}
        {\includegraphics[width=0.8\textwidth]{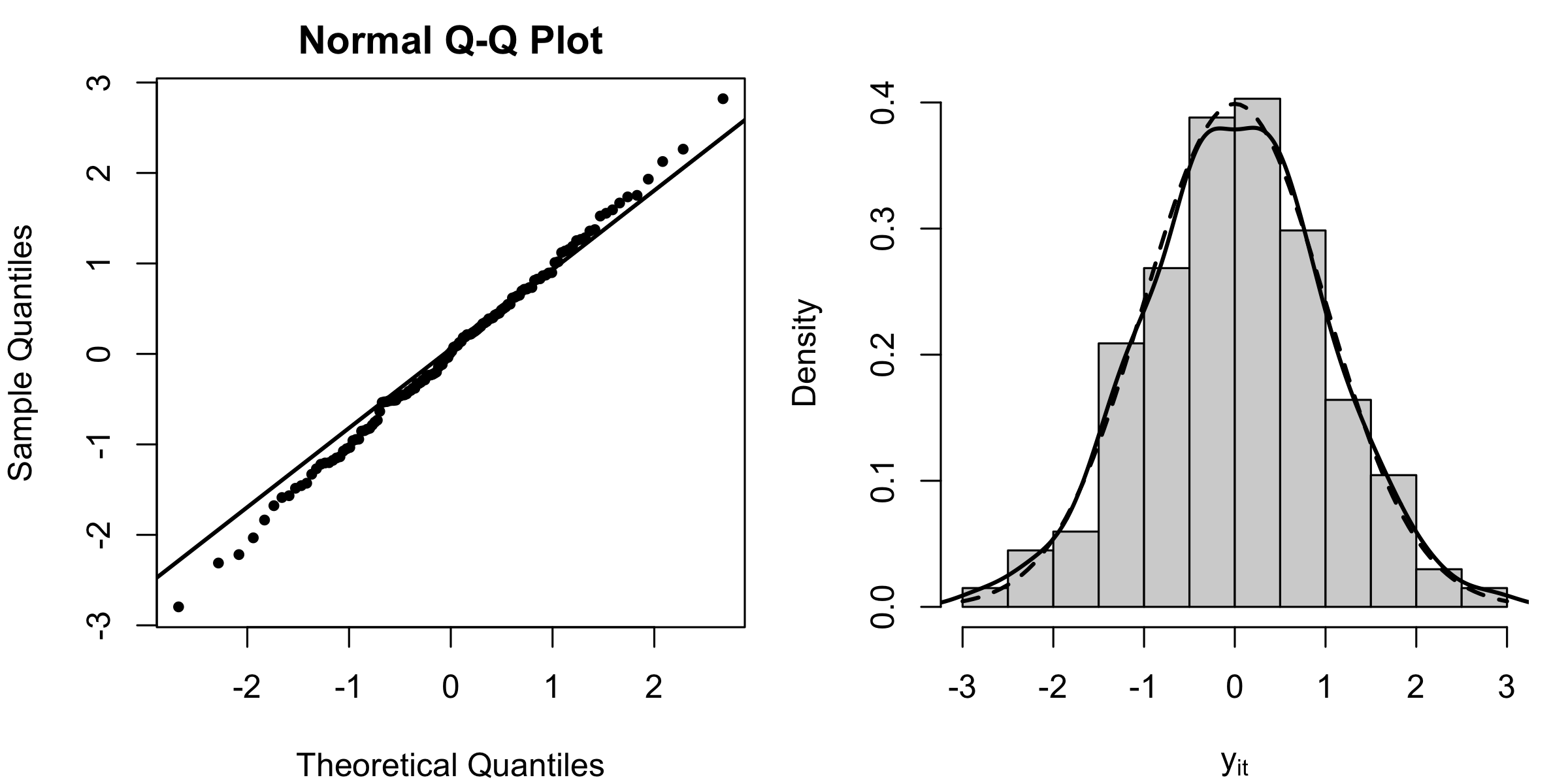}}
    \end{center}
    \caption{\label{appendix:fig:adni_brain}Normal Q-Q plot and histogram of the fMRI data extracted from one brain region of one subject in the ADNI study.}
\end{figure}

\clearpage

\bibliographystyle{apalike}
\bibliography{bibliography}

@article{khalili2010new,
  title={New estimation and feature selection methods in mixture-of-experts models},
  author={Khalili, Abbas},
  journal={Canadian Journal of Statistics},
  volume={38},
  number={4},
  pages={519--539},
  year={2010},
  publisher={Wiley Online Library}
}

@article{cavedo2018sex,
  title={Sex differences in functional and molecular neuroimaging biomarkers of {Alzheimer's} disease in cognitively normal older adults with subjective memory complaints},
  author={Cavedo, Enrica and Chiesa, Patrizia A and Houot, Marion and Ferretti, Maria Teresa and Grothe, Michel J and Teipel, Stefan J and Lista, Simone and Habert, Marie-Odile and Potier, Marie-Claude and Dubois, Bruno and others},
  journal={Alzheimer's \& Dementia},
  volume={14},
  number={9},
  pages={1204--1215},
  year={2018},
  publisher={Elsevier}
}

@article{li2021sex,
  title={Sex Difference in Network Topology and Education Correlated With Sex Difference in Cognition During the Disease Process of {Alzheimer}},
  author={Li, Xiaoshu and Zhou, ShanShan and Zhu, Wanqiu and Li, Xiaohu and Gao, Ziwen and Li, Meiqin and Luo, Shilei and Wu, Xingqi and Tian, Yanghua and Yu, Yongqiang},
  journal={Frontiers in Aging Neuroscience},
  volume={13},
  pages={241},
  year={2021},
  publisher={Frontiers}
}

@article{panda2014unraveling,
  title={Unraveling Brain Functional Connectivity of encoding and retrieval in the context of education},
  author={Panda, Rajanikant and Bharath, Rose Dawn and George, Lija and Kanungo, Silpa and Reddy, Rajakumari P and Upadhyay, Neeraj and Thamodharan, Arumugam and Rajeshwaran, Jamuna and Rao, Shobini L and Gupta, Arun Kumar},
  journal={Brain and Cognition},
  volume={86},
  pages={75--81},
  year={2014},
  publisher={Elsevier}
}

@article{perry2017independent,
  title={The independent influences of age and education on functional brain networks and cognition in healthy older adults},
  author={Perry, Alistair and Wen, Wei and Kochan, Nicole A and Thalamuthu, Anbupalam and Sachdev, Perminder S and Breakspear, Michael},
  journal={Human Brain Mapping},
  volume={38},
  number={10},
  pages={5094--5114},
  year={2017},
  publisher={Wiley Online Library}
}

@Article{tibshirani2005sparsity,
  author    = {Tibshirani, Robert and Saunders, Michael and Rosset, Saharon and Zhu, Ji and Knight, Keith},
  title     = {Sparsity and smoothness via the fused lasso},
  journal   = {Journal of the Royal Statistical Society: Series B (Statistical Methodology)},
  year      = {2005},
  volume    = {67},
  number    = {1},
  pages     = {91--108},
  publisher = {Wiley Online Library},
}

@article{smith2004advances,
  title={Advances in functional and structural {MR} image analysis and implementation as {FSL}},
  author={Smith, Stephen M and Jenkinson, Mark and Woolrich, Mark W and Beckmann, Christian F and Behrens, Timothy EJ and Johansen-Berg, Heidi and Bannister, Peter R and De Luca, Marilena and Drobnjak, Ivana and Flitney, David E and others},
  journal={NeuroImage},
  volume={23},
  pages={S208--S219},
  year={2004},
  publisher={Elsevier}
}

@article{ieva2016,
  author  = {Francesca Ieva and Anna Maria Paganoni and Nicholas Tarabelloni},
  title   = {Covariance-based Clustering in Multivariate and Functional Data Analysis},
  journal = {Journal of Machine Learning Research},
  year    = {2016},
  volume  = {17},
  number  = {143},
  pages   = {1--21},
  url     = {http://jmlr.org/papers/v17/15-568.html}
}

@article{banfield1993model,
  title={Model-based {Gaussian} and non-{Gaussian} clustering},
  author={Banfield, Jeffrey D and Raftery, Adrian E},
  journal={Biometrics},
  pages={803--821},
  year={1993},
  publisher={JSTOR}
}

@article{celeux1995gaussian,
  title={Gaussian parsimonious clustering models},
  author={Celeux, Gilles and Govaert, G{\'e}rard},
  journal={Pattern Recognition},
  volume={28},
  number={5},
  pages={781--793},
  year={1995},
  publisher={Elsevier}
}

@article{dempster1977maximum,
  title={Maximum likelihood from incomplete data via the {EM} algorithm},
  author={Dempster, Arthur P and Laird, Nan M and Rubin, Donald B},
  journal={Journal of the Royal Statistical Society: Series B (Methodological)},
  volume={39},
  number={1},
  pages={1--22},
  year={1977},
  publisher={Wiley Online Library}
}

@article{jacobs1991adaptive,
  title={Adaptive mixtures of local experts},
  author={Jacobs, Robert A and Jordan, Michael I and Nowlan, Steven J and Hinton, Geoffrey E},
  journal={Neural Computation},
  volume={3},
  number={1},
  pages={79--87},
  year={1991},
  publisher={MIT Press}
}

@article{murphy2020gaussian,
  title={Gaussian parsimonious clustering models with covariates and a noise component},
  author={Murphy, Keefe and Murphy, Thomas Brendan},
  journal={Advances in Data Analysis and Classification},
  volume={14},
  number={2},
  pages={293--325},
  year={2020},
  publisher={Springer}
}

@article{dilernia2022penalized,
  title={Penalized model-based clustering of {fMRI} data},
  author={Dilernia, Andrew and Quevedo, Karina and Camchong, Jazmin and Lim, Kelvin and Pan, Wei and Zhang, Lin},
  journal={Biostatistics},
  volume={23},
  number={3},
  pages={825--843},
  year={2022},
  publisher={Oxford University Press}
}

@article{liu2023simultaneous,
  title={Simultaneous cluster structure learning and estimation of heterogeneous graphs for matrix-variate {fMRI} data},
  author={Liu, Dong and Zhao, Changwei and He, Yong and Liu, Lei and Guo, Ying and Zhang, Xinsheng},
  journal={Biometrics},
  volume={79},
  number={3},
  pages={2246--2259},
  year={2023},
  publisher={Oxford University Press}
}

@article{zhao2021covariate,
  title={Covariate assisted principal regression for covariance matrix outcomes},
  author={Zhao, Yi and Wang, Bingkai and Mostofsky, Stewart H and Caffo, Brian S and Luo, XI},
  journal={Biostatistics},
  volume={22},
  number={3},
  pages={629--645},
  year={2021},
  publisher={Oxford University Press}
}

@article{zeng2014unsupervised,
  title={Unsupervised classification of major depression using functional connectivity {MRI}},
  author={Zeng, Ling-Li and Shen, Hui and Liu, Li and Hu, Dewen},
  journal={Human brain mapping},
  volume={35},
  number={4},
  pages={1630--1641},
  year={2014},
  publisher={Wiley Online Library}
}

@article{yang2022study,
  title={A study of brain networks for autism spectrum disorder classification using resting-state functional connectivity},
  author={Yang, Xin and Zhang, Ning and Schrader, Paul},
  journal={Machine Learning with Applications},
  volume={8},
  pages={100290},
  year={2022},
  publisher={Elsevier}
}

@article{dennis2014functional,
  title={Functional brain connectivity using {fMRI} in aging and {Alzheimer's} disease},
  author={Dennis, Emily L and Thompson, Paul M},
  journal={Neuropsychology Review},
  volume={24},
  pages={49--62},
  year={2014},
  publisher={Springer}
}

@article{clementz2016identification,
  title={Identification of distinct psychosis biotypes using brain-based biomarkers},
  author={Clementz, Brett A and Sweeney, John A and Hamm, Jordan P and Ivleva, Elena I and Ethridge, Lauren E and Pearlson, Godfrey D and Keshavan, Matcheri S and Tamminga, Carol A},
  journal={American Journal of Psychiatry},
  volume={173},
  number={4},
  pages={373--384},
  year={2016},
  publisher={Am Psychiatric Assoc}
}

@article{yamada2017resting,
  title={Resting-state functional connectivity-based biomarkers and functional MRI-based neurofeedback for psychiatric disorders: a challenge for developing theranostic biomarkers},
  author={Yamada, Takashi and Hashimoto, Ryu-ichiro and Yahata, Noriaki and Ichikawa, Naho and Yoshihara, Yujiro and Okamoto, Yasumasa and Kato, Nobumasa and Takahashi, Hidehiko and Kawato, Mitsuo},
  journal={International Journal of Neuropsychopharmacology},
  volume={20},
  number={10},
  pages={769--781},
  year={2017},
  publisher={Oxford University Press US}
}

@article{schwarz1978estimating,
  title={Estimating the dimension of a model},
  author={Schwarz, Gideon},
  journal={The Annals of Statistics},
  pages={461--464},
  year={1978},
  publisher={JSTOR}
}

@article{friston2011functional,
  title={Functional and effective connectivity: a review},
  author={Friston, Karl J},
  journal={Brain Connectivity},
  volume={1},
  number={1},
  pages={13--36},
  year={2011},
  publisher={Mary Ann Liebert, Inc. 140 Huguenot Street, 3rd Floor New Rochelle, NY 10801 USA}
}

@article{lin2024latent,
  title={Latent subgroup identification in image-on-scalar regression},
  author={Lin, Zikai and Si, Yajuan and Kang, Jian},
  journal={The Annals of Applied Statistics},
  volume={18},
  number={1},
  pages={468},
  year={2024},
  publisher={NIH Public Access}
}

@article{chen2022simultaneous,
  title={Simultaneous differential network analysis and classification for matrix-variate data with application to brain connectivity},
  author={Chen, Hao and Guo, Ying and He, Yong and Ji, Jiadong and Liu, Lei and Shi, Yufeng and Wang, Yikai and Yu, Long and Zhang, Xinsheng and Alzheimers Disease Neuroimaging Initiative},
  journal={Biostatistics},
  volume={23},
  number={3},
  pages={967--989},
  year={2022},
  publisher={Oxford University Press}
}

@article{lee2015geodesic,
  title={Geodesic clustering for covariance matrices},
  author={Lee, Haesung and Ahn, Hyun-Jung and Kim, Kwang-Rae and Kim, Peter T and Koo, Ja-Yong},
  journal={Communications for Statistical Applications and Methods},
  volume={22},
  number={4},
  pages={321--331},
  year={2015},
  publisher={The Korean Statistical Society}
}

@article{verdinelli2019hybrid,
  title={Hybrid {Wasserstein} distance and fast distribution clustering},
  author={Verdinelli, Isabella and Wasserman, Larry},
  year={2019}
}

\end{document}